\documentclass[journal]{IEEEtran}
\usepackage[utf8]{inputenc} 
\usepackage[T1]{fontenc}
\usepackage{amsfonts}
\usepackage[cmex10]{amsmath} 
\usepackage{url}
\usepackage{lettrine}
\usepackage{framed}
\usepackage{color}
\usepackage{algorithm}
\usepackage{algorithmic}
\usepackage{overpic}
\usepackage{dsfont}
\usepackage{booktabs} 
\usepackage{framed}
\usepackage{bm}
\usepackage{bbm}
\usepackage{ifthen}
\usepackage{cite}
\usepackage{graphicx}
\usepackage{dsfont}
\usepackage{framed}
\usepackage{nicefrac}
\usepackage{bm}
\usepackage{overpic}
\usepackage{xcolor}
\usepackage{MnSymbol}
\usepackage{wasysym}
\usepackage{balance}

\newtheorem{theorem}{Theorem}
\newtheorem{proposition}{Proposition}

\usepackage{xr}

\newcommand{\edit}[1]{{\color{black} #1}}
\newcommand{\editr}[1]{{\color{black} #1}}

\ifCLASSINFOpdf
\else
\fi
\begin{document}
%
\title{Optimizing Age-of-Information in Adversarial and Stochastic Environments}
%
%
%

       
\author{
\IEEEauthorblockN{Abhishek Sinha\IEEEauthorrefmark{1} and Rajarshi Bhattacharjee\IEEEauthorrefmark{2}\\}
\IEEEauthorblockA{\IEEEauthorrefmark{1}School of Technology and Computer Science,  Tata Institute of Fundamental Research, Mumbai 400 005, India\\
\IEEEauthorrefmark{2} College of Information and Computer Sciences, University of Massachusetts at Amherst\\}
Email: 
\IEEEauthorrefmark{1}abhishek.sinha@tifr.res.in,
\IEEEauthorrefmark{2}rbhattacharj@cs.umass.edu

\thanks{This paper was presented in part at  \cite{srivastava2019minimizing}, \cite{banerjee2020fundamental}, and \cite{bhattacharjee2020competitive}. The work by Rajarshi Bhattacharjee was completed while he was working with the first author as a project associate at the Indian Institute of Technology Madras.}
}
\maketitle

\begin{abstract}
We design efficient online scheduling policies to maximize the freshness of information delivered to the users in a cellular network under both adversarial and stochastic channel and mobility assumptions. The information freshness achieved by a policy is investigated through the lens of a recently proposed metric - \emph{Age-of-Information} (AoI). We show that a natural greedy scheduling policy is competitive against any optimal offline policy in minimizing the AoI in the adversarial setting. We also derive universal lower bounds to the competitive ratio achievable by any online policy in the adversarial framework.  In the stochastic setting, we show that a simple index policy is near-optimal for minimizing the average AoI in two different mobility scenarios. Further, we prove that the greedy scheduling policy minimizes the peak AoI for static users in the stochastic setting. Simulation results show that the proposed policies perform well under realistic assumptions.    
\end{abstract}


%
\IEEEpeerreviewmaketitle

\section{Introduction}
\IEEEPARstart{Q}{uality}-of-Service (QoS) offered by data networks has been traditionally evaluated along three major  dimensions, namely, throughput, latency, and energy/spectral efficiency. There is an extensive body of literature on cross-layer resource allocation algorithms to optimize the above metrics in various wired and wireless networks \cite{tassiulas, mandelbaum2004scheduling, sinha_umw, neely2010stochastic, kozat2004framework}. However, it has been widely argued that the standard QoS metrics are primarily geared towards quantifying the degree of utilization of system resources and less towards measuring the actual user experience \cite{new_QoS}. With the explosive growth of hand-held mobile devices and the advent of the Internet of Things (IoT) and real-time AR and VR systems, future networks aim to optimize the Quality of Experience (QoE) for its end-users 
\cite{QoE}. In order to incorporate the QoE objectives directly into the design decisions, a fundamentally new metric, called \emph{Age-of-Information} (AoI), has been proposed recently for measuring the freshness of information available to the users \cite{kaul2012real, kosta2017age}. Informally, the AoI of a user is defined as the average length of time between its successive information updates. From the perspective of a network operator, minimizing either the average or the peak AoI of the users may be of interest \cite{farazi2019fundamental}. For example, in the case of non-critical status updates, it makes sense to minimize the average AoI of all users. On the other hand, in applications where the most outdated node is the bottleneck (\emph{e.g.,} mission critical or industrial IoT applications), minimizing the peak AoI across all devices is a reasonable objective. Apart from uncertain wireless channels and bandwidth constraints, the mobility of the users poses significant challenges to optimize the AoI. It is well-known that mobility increases the capacity of wireless ad hoc networks \cite{grossglauser2002mobility, gamal2004throughput, neely2005capacity}. \edit{Furthermore, the effect of mobility on the QoS, latency, and capacity of wireless cellular networks have been investigated in the papers \cite{baynat2015impact, anton2019impact, baiocchi1996effects}.} However, to the best of our knowledge, the effect of user mobility on the freshness of information has not been previously explored. In this paper, we design near-optimal scheduling policies for minimizing the average and peak AoI in a cellular network under two widely different channel and mobility scenarios. 
   
For analytical tractability, most of the existing papers on wireless communication work with stationary channel models \cite{hsu2007modeling}. In rapidly varying environments, such as high-speed trains and vehicle-to-vehicle communication, the stationarity assumption no longer holds in practice. This is particularly true with the emerging 5G mmWave technology, which needs to accurately beamform the wireless signal to mitigate severe attenuation loss at high frequencies \cite{non_stationary, wu2017general}. On the other hand, designing accurate and analytically tractable non-stationary channel models has remained an overarching challenge to the research community for decades \cite{nonstat1, nonstat2}. Furthermore, accurate channel estimation in rapidly-varying environments is often infeasible for applications requiring ultra-low latency. Responding to this challenge, we investigate the following question in the first half of the paper - is it possible to design a scheduling policy that minimizes the AoI irrespective of the channel dynamics and user-mobility patterns? The question is considerably general, as we do not make any assumption on either the channel statistics or the user-mobility, both of which may be dictated by an omniscient adversary in the worst case. The adversarial model is also useful for guaranteeing reliable communication in the presence of tactical jammers, where the interferers, in reality, may behave adversarially \cite{poisel2011modern, mpitziopoulos2009survey}. A similar problem in the context of \emph{stability} of wireless networks was considered in \cite{andrews2007routing} under adversarial arrival and link transmission rates.

To formalize the problem, we first introduce an adversarial binary erasure channel model, which may be considered as an adversarial counterpart of the celebrated Binary Erasure Channel (\textsf{BEC}) model. However, unlike a similar adversarial model considered in \cite{bassily2014causal}, we neither make any causality assumption nor impose any upper-bound on the fraction of erasures. Thus, our channel model is considerably more general. We propose a greedy scheduling policy that competitively minimizes the average and peak AoI in this model. See Table \ref{summary} for our main results for the adversarial framework.  

In contrast with the non-stationary environment, minimizing the AoI in stationary stochastic environments has been extensively studied. In the paper \cite{kadota2018scheduling}, the authors study the average AoI minimization problem for static users associated with a single access point. The authors show that the greedy Max-Age (\textsf{MA}) policy is optimal for minimizing the average AoI in a symmetric static network. In the same paper, the authors also propose a $4$-optimal Max-Weight scheduling policy (\textsf{MW}) for an arbitrary static network consisting of a single access point. The proposed \textsf{MW} policy has been reported to practically improve the information freshness in WiFi networks by two orders of magnitude \cite{IgorMobicom}. The paper \cite{kadota2018optimizing} extends the \textsf{MW} policy by taking into account additional throughput constraints. The paper \cite{talak2017minimizing} designs optimal stationary scheduling policies for minimizing the AoI in multi-hop networks with static users under general interference constraints. The paper \cite{tripathi2019age} considers the problem of designing an AoI-optimal trajectory for a mobile agent to facilitate the information dissemination from a central station to a set of ground terminals. However, the problem of designing an AoI optimal scheduling policy for mobile users has not been considered before. In the second half of the paper we tackle this question and show that a natural multi-cell extension of the Max-Weight policy performs well under certain mobility assumptions.


\begin{table*}
\caption{Summary of the results on the Competitive Ratios in the adversarial framework} \label{summary}
\centering
\begin{tabular}{ccccccc}
\toprule
\textbf{Metrics} & \textbf{\textsf{Cost} function} & \textbf{Mobility} & \textbf{Upper Bound} & \textbf{Achieving Policy}  & \textbf{Lower Bound} & \textbf{Optimality gap} \\
\midrule
Average AoI & $N^{-1}\sum_{t=1}^T\sum_{i=1}^Nh_i(t)$        &\textsf{Yes} & $O(N^2)$ & \textsf{CMA} & $O(N)$ & $O(N)$   \\
\midrule
Peak AoI  & $\sum_{t=1}^T\max_{i=1}^N h_i(t)$  & \textsf{Yes} & $O(N)$ & \textsf{CMA} & $\Omega(\frac{N}{\ln(N)})$ & $O(\ln(N))$\\
\bottomrule
\end{tabular}
\end{table*} 
\subsection*{Our contributions:}
 In this paper, we consider the AoI-optimal scheduling problem for mobile users in both adversarial and stochastic models. Our objective is to design simple scheduling policies that performs competitively in the adversarial environments and achieves near-optimality in the more benign stationary stochastic environments. The adversarial formulation of the problem is new and, to the best of our knowledge, has not been considered before. On the other hand, while the AoI-optimal scheduling problem for static users in the stochastic setting is now well-understood \cite{kadota2018optimizing, kadota2018scheduling}, the multi-cell extension of the problem with mobile users is new.
 In particular, we make the following contributions in this paper:\\
\begin{enumerate}
\item Within the adversarial framework of Section \ref{adversarial}, we show that a greedy online scheduling policy is $2N^2$-competitive for the Average AoI metric. Using Yao's minimax principle, we also establish a fundamental lower bound by showing that no online policy can have a competitive ratio smaller than $2N-1$. 
\item For the peak AoI metric, we show that the same greedy scheduling policy achieves a competitive ratio of $2N$ in the adversarial setting. Using Yao's minimax principle, we show that no online policy can have a competitive ratio better than $\Omega(\frac{N}{\ln(N)})$. Table \ref{summary} summarizes our main results in the adversarial model.
\item In Section \ref{stochastic}, we consider the AoI-optimal multi-user scheduling problem for mobile users in a stationary environment. For the average-age metric, we design a $2$-optimal scheduling policy for mobile users with i.i.d. uniform mobility. As a by-product of our analysis, we improve upon the best-known $4$-approximation bound known for static users \cite{kadota2018scheduling, kadota2019scheduling}. 
\item To minimize the peak AoI metric in the stochastic setting, we show that the greedy policy is optimal for a single-cell static network. We give a short proof of this optimality result by exhibiting a solution to a countable-state average-cost MDP problem in the ``closed-form", which might be of independent interest. This result supplements Theorem 5 of \cite{kadota2018scheduling}, which establishes the optimality of the greedy policy for the average AoI metric for \emph{symmetric} static networks using stochastic dominance arguments. We further show that the greedy policy achieves the optimal large-deviation rate. Table \ref{stoch-table} summarizes our main results in the stochastic model.
\end{enumerate} 
\edit{
\paragraph*{Discussion}
Our motivation for studying the adversarial and the stochastic models together stems from the following two reasons - (1) to highlight the complementary strengths and weaknesses of each model, and (2) to show that, despite the differences, there exist simple scheduling policies that perform well in both settings. More specifically, 
 \begin{enumerate}
\item  While the stochastic model makes strong assumptions on the environment, the adversarial framework makes virtually no assumptions. Hence, the adversarial model allows us to study the optimal scheduling problem in an arbitrary non-stationary environment, for which very few results exist in the literature. However, because of the difference in the generality of the assumptions, considerably stronger performance guarantees can be established in the stochastic setting compared to the adversarial environment. This brings us to the next point.

\item Given the difference in the performance guarantees, it is natural to ask whether there exist ``universal" policies that perform near-optimally in the stochastic setting and simultaneously enjoy non-trivial performance guarantees in the adversarial environment. We answer the above question in the affirmative by exhibiting a simple greedy scheduling policy that achieves the above goal. 
\end{enumerate}

}

The rest of the paper is organized as follows. In Section \ref{adversarial}, we describe the adversarial model and prove upper and lower bounds on the competitive ratio achievable within this framework for minimizing AoI. In Section \ref{stochastic}, we describe the stochastic model and design near-optimal policies for minimizing the AoI in the stochastic setting. In Section \ref{simulation}, we compare the performance of the proposed scheduling policies via numerical simulations. Section \ref{conclusion} concludes the paper with some pointers to a few related open problems.

\section{AoI Minimization in Adversarial Environments} \label{adversarial} 

We consider an optimal downlink scheduling problem in a wireless cellular network where $N$ users roam around in a region having $M$ Base Stations (BS). The environment, which is entirely specified by the channel states and the user mobility pattern, can evolve in an arbitrary fashion. Instead of trying to fit a complicated probabilistic model with multiple parameters \cite{nonstat1, nonstat2}, we take a conservative view and model the environment using an adversarial framework. The adversarial viewpoint can be practically motivated by considering URLLC-type traffic, which requires extremely low latency with very high reliability \cite{urllc}. Besides being analytically tractable, all achievability results in the adversarial model (Theorem \ref{max_upper_bound}) carry over to more benign stochastic environments. Moreover, as we will see in the sequel, policies having a good competitive ratio in the adversarial setting sometimes translate to optimal policies in the stochastic setting (Theorem \ref{opt}). 
\begin{figure}
\centering
\begin{overpic}[width=0.35\textwidth]{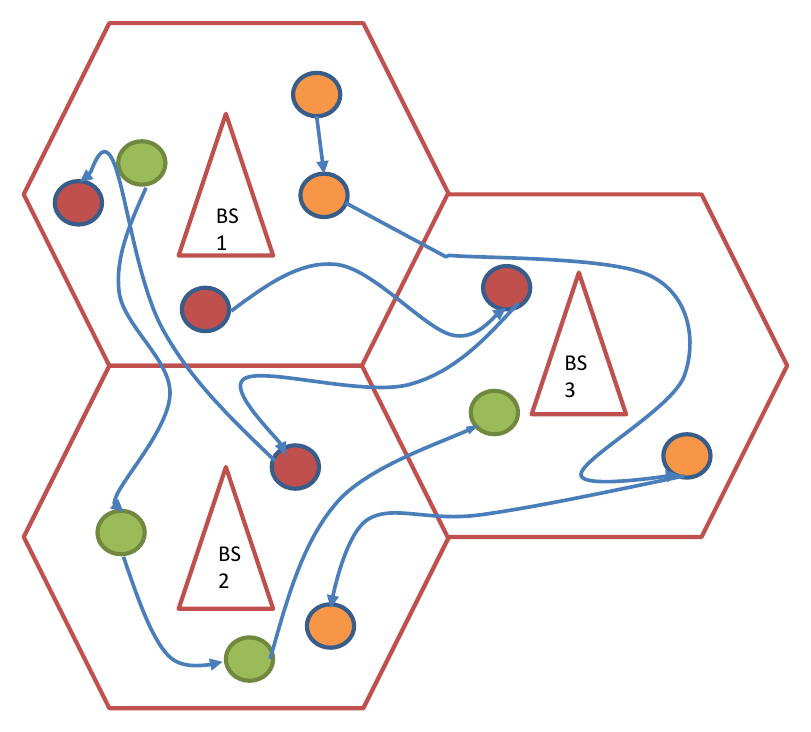}
\put(28,75.5){\color{blue} \thicklines \vector(-1,-4){5.5}}
\put(28,32){\color{blue} \thicklines \vector(-4,-3){12}}
\put(71.5,55.5){\color{blue} \thicklines \vector(3,-4){15}}
\put(80, 45){\footnotesize{transmission}}
\end{overpic}
\caption{Illustrating the movement of $N=3$ users (each with a distinct color) in an area with $M=3$ cells. The blue arrows indicate packet transmissions by the BS.}
\label{AoI_mobility_fig}
\end{figure}
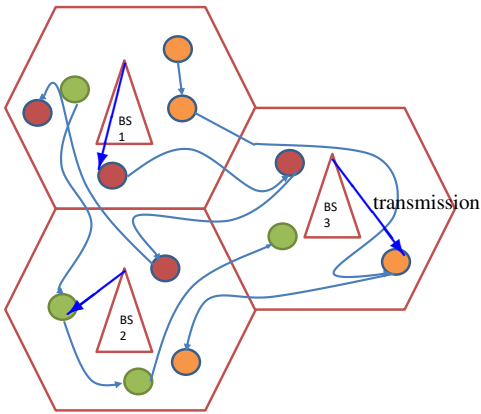
\subsection{Adversarial System Model} 
The main system components are described below.
\paragraph{Network model} The area covered by a Base Station (henceforth referred to as \emph{BS}) is referred to as a \emph{cell}. The cells are assumed to be spatially disjoint. Time is slotted, and at each time slot, a user can either stay in its current cell or move to any other $M-1$ cells (the new cell need not be adjacent). The movement could be dictated by an omniscient adversary. Our mobility model is considerably general, as it does not make any assumptions (statistical or otherwise) on the speed or user movement patterns. See Figure \ref{AoI_mobility_fig} for a schematic. 
\paragraph{Traffic Model} We consider a \emph{saturated} traffic model where at the beginning of every slot, each of the $M$ Base Stations receives a fresh update packet for each user from an external source (\emph{e.g.,} a high-speed optical core network). Since our objective is to maximize the freshness of information at the user-end, any stale packet at the BS buffers is replaced by incoming fresh packets at each slot. \edit{Recall that a stale packet for a user at a slot is defined as any buffered packet(s) from the previous slot.}  Each BS can beamform and schedule a downlink packet transmission at each slot to only one user within its coverage area. \edit{Although to be specific, we consider downlink scheduling in this paper, an analogous problem can also be formulated for uplink transmissions, and all of our results apply to this case as well.}

The saturated traffic model is standard in applications which involve continuous status updates \cite{costa2016age}, such as monitoring and surveillance with sensor networks \cite{javani2019age}, velocity and position updates for autonomous vehicles \cite{kaul2011minimizing}, command and control information exchange in mission-critical systems, disseminating stock-index updates and live game scores. \edit{The saturated traffic model has also been used earlier in connection with designing rate-optimal scheduling policies \cite{borst2001dynamic, kushner2004convergence, holtzman2001asymptotic}. The advantage of this model is that it separates the arrival process from the scheduling policy, leading to a cleaner analysis in both adversarial and stochastic settings.}
\paragraph{Channel states, Control, and Objective} The policy controls and optimizes  user scheduling decisions by each BS. We consider a binary erasure channel model where the channel state for any user at any slot can be either \textsf{Good} or \textsf{Bad} (also referred to as \textsf{ON/OFF} in the literature \cite{eytan-on-off, ying2006large}). \edit{Here a \emph{slot} is defined to be the duration of the entire packet transmission. In practice (\emph{e.g.,} in 5G systems), a packet is transmitted over several resource blocks (RBs) consisting of multiple OFDM symbols. Thus, if the aggregate of channel states remains poor over a packet transmission block so that the probability of a packet decoding error remains sufficiently large (\emph{e.g.,} an outage event), the channel for that slot is considered to be in \textsf{Bad} (or \textsf{OFF}) state. Else, the channel state for that slot is considered to be \textsf{Good} (or \textsf{ON}).}

\begin{figure}
\centering
\begin{overpic}[width=0.5\textwidth]{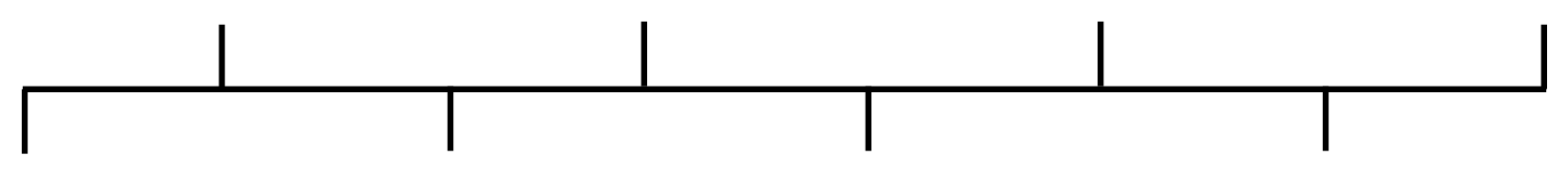}
\end{overpic}
\put(-273,-5){\footnotesize{Beginning of}}
\put(-260,-15){\footnotesize{Slot}}
\put(-255,30){\footnotesize{User movements}}
\put(-210,-5){\footnotesize{Fresh Packet}}
\put(-212,-15){\footnotesize{arrivals at BS}}
\put(-180,30){\footnotesize{User scheduling}}
\put(-145,-5){\footnotesize{Revelation of the}}
\put(-150,-15){\footnotesize{channel states }(\textsf{G}/\textsf{B})}
\put(-110,40){\footnotesize{Success/Failure}}
\put(-115,30){\footnotesize{of the transmissions}}
\put(-70,-5){\footnotesize{Measurement of}}
\put(-55,-15){\footnotesize{age $\bm{h}(t)$}}
\put(-35,30){\footnotesize{End of Slot}}
\vspace{10pt}
\caption{Event sequence at each slot}
\label{event_sequence}
\end{figure}
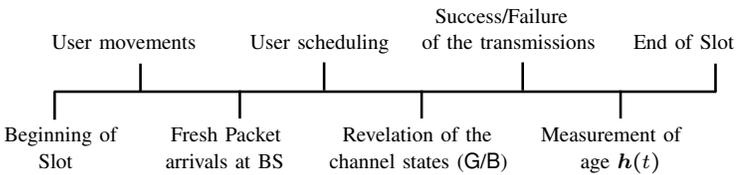

The schedulers are assumed to be oblivious to the current channel states (\emph{i.e.,} no \textsc{CSIT}). An online scheduling policy $\pi$ first selects a user in each cell (if the cell contains at least one user), and then transmits the latest packet from the BS to the selected users over the wireless channel. If the corresponding channel is in \textsf{Good} state, the user decodes the packet successfully. Otherwise, the packet is lost. A lost packet is never retransmitted as the scheduler receives fresh packets at every slot. In the adversarial model, we posit that the channel states are dictated by an omniscient adversary \cite{andrews2007routing}. In particular, we allow the situation where the adversary knows the scheduling policy in advance and chooses the channel realizations \emph{after} the scheduling decisions have been made for a slot. On the other hand, the scheduling policy $\pi$ is necessarily \emph{online} and has no information about the channel states  in the current or future slots. The set of all admissible scheduling policies is denoted by $\Pi.$ See Figure \ref{event_sequence} for the timeline of events taking place at every slot.

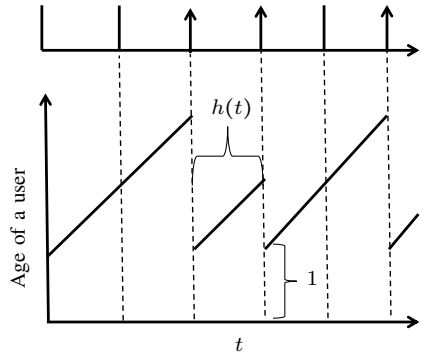
\begin{figure}[H]
\centering
\begin{overpic}[width=0.3\textwidth]{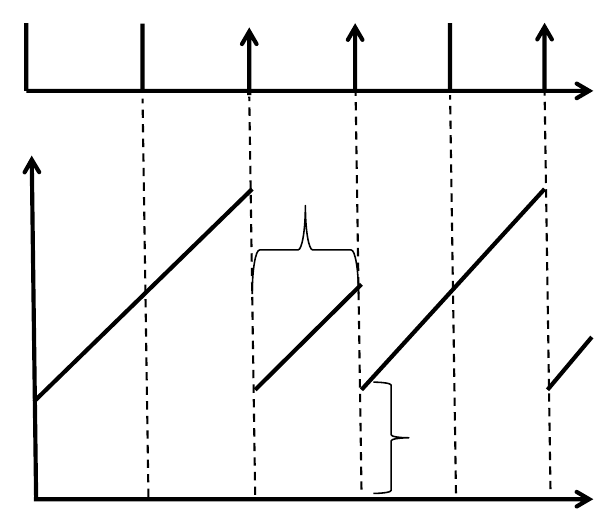}
\end{overpic}
\put(-160,21){\rotatebox{90}{\footnotesize{Age of a user}}}
\put(-84,87){\footnotesize{$h(t)$}}
\put(-48,22){\footnotesize{$1$}}
\put(-75,-3){\footnotesize{$t$}}
\caption{Time-evolution of the Age-of-Information for a user. Solid arrows denote successful transmissions.}
\label{AoI_fig}
\end{figure}

We are interested in competitively optimizing the information freshness for all users. Formally, our objective is to design a decentralized scheduling policy that minimizes some measure of the aggregate Age-of-Information of the users as defined next. For any slot $t \geq 1$, let $t_i(t) < t$ denote the last time prior to time $t$ at which the $i$\textsuperscript{th} user successfully received a packet from some BS. The Age-of-Information (AoI) of the user at time $t$, denoted by $h_i(t)$, is defined as: 
\[ h_i(t) \equiv t-t_i(t). \]
In other words, $h_i(t)$ denotes the length of the time elapsed since the $i$\textsuperscript{th} user received its last update packet before time $t$. Thus, the process $\{h_i(t)\}_{t \geq 1}$ quantifies the \emph{staleness} (or age) of the information available to the $i$\textsuperscript{th} user. Accordingly, we define an $N$-dimensional state-vector $\bm{h}(t)$, whose $i$\textsuperscript{th} component, $h_i(t),$ denotes the AoI of the $i$\textsuperscript{th} user at time $t$. Clearly, the plot of $h_i(t)$ vs. time $t$ has a saw-tooth shape that increases linearly with unit-slope until a fresh packet is received. Upon the reception of a fresh packet, the AoI $h_i(t)$ instantaneously drops to $1$. From that point onwards, $h_i(t)$ again increases linearly, repeating the saw-tooth pattern \cite{kadota2018scheduling}.
See Figure \ref{AoI_fig} for an illustration. In this paper, we consider optimizing the following two different aggregate AoI metrics:
\paragraph*{a) \textbf{Average AoI}} The time-averaged cost corresponding to the average AoI for $N$ users up to time $T$ is defined as:
\begin{eqnarray}\label{avg_cost_fn}
\textsf{AoI}_{\textrm{avg}}(T) = \frac{1}{NT}\sum_{t=1}^{T}\bigg(\sum_{i=1}^N h_i(t)\bigg). 
\end{eqnarray}
\paragraph*{b) \textbf{Peak AoI}} The instantaneous peak AoI at a slot is defined as the maximum age among all users. The time-averaged cost corresponding to the peak AoI for a time-horizon of length $T$ is defined as:
\begin{eqnarray}\label{max_cost_fn}
\textsf{AoI}_{\textrm{peak}}(T) =\frac{1}{T}\sum_{t=1}^{T}\max_{i=1}^{N} h_i(t)
\end{eqnarray}

\paragraph*{\textbf{Performance Metric}}
As standard in the literature on online algorithms \cite{fiat1998online, albers1996competitive}, we gauge the performance of any online scheduling policy $\mathcal A$ using the notion of \emph{competitive ratio} (denoted by $\eta^{\mathcal A}$). Recall that the competitive ratio is defined as the worst-case ratio of the cost incurred by an online policy to that of an optimal \emph{offline} policy \textsf{OPT}. The \textsf{OPT} policy is assumed to be equipped with hindsight knowledge. Specializing to our context, let $\bm{\sigma} $ be a sequence representing the channel states and the user's locations for the entire time-horizon. Then, the competitive ratio of a policy $\mathcal A$ is defined as \cite{albers1996competitive}:
\begin{eqnarray}\label{comp_rat_def}
\eta^{\mathcal{A}} = \sup_{\bm \sigma}\bigg(\frac{\textrm{Cost of the policy } \mathcal A \textrm{ on } \bm{\sigma}}{\textrm{Cost of OPT on } \bm{\sigma}}\bigg).	
\end{eqnarray}
In the above definition, the supremum is taken over \emph{all} finite-length admissible sequences $\bm \sigma$. Depending on the objective, the cost function in the definition \eqref{comp_rat_def} can be taken to be either Eqn.\ \eqref{avg_cost_fn} or Eqn.\ \eqref{max_cost_fn}. We emphasize that, while the online policy $\mathcal A$ has only causal information (\emph{i.e.,} knows only the subsequence $\bm{\sigma}_{1}^{t-1}$ at time $t$), the policy \textsf{OPT} is assumed to be equipped with non-causal knowledge of the entire sequence $\bm \sigma_{1}^T$ right at the beginning. Our objective is to design an online scheduling policy $\mathcal{A}$ with a small competitive ratio so that it performs close to the \textsf{OPT} policy. 
 \paragraph*{Discussion} 
 In the sum-rate maximization problem, the objective is to maximize the total rate of successful packet transmissions to the users \cite{tassiulas}. 
 It is interesting to note that finding a scheduling policy with a small competitive ratio for the sum-throughput objective is too strong a requirement, as all deterministic policies suffer from unbounded competitive ratios.
 This can be understood from the following simple example. Assume that two stationary users are located in a single cell. If an online policy $\mathcal{A}$ schedules the first user at any slot, the adversary can set the channel corresponding to the first user to \textsf{Bad} and the second user's channel to \textsf{Good} and vice versa. Hence, all transmissions by policy $\mathcal{A}$ are unsuccessful. On the other hand, at any slot, the optimal policy schedules the user that has the \textsf{Good} channel state for that slot. Hence, the \textsf{OPT} policy achieves unit throughput, resulting in an unbounded competitive ratio. In the following, we show that, somewhat surprisingly, there exists a policy that achieves a finite competitive ratio for the AoI minimization problem.
\subsection{Achievability}
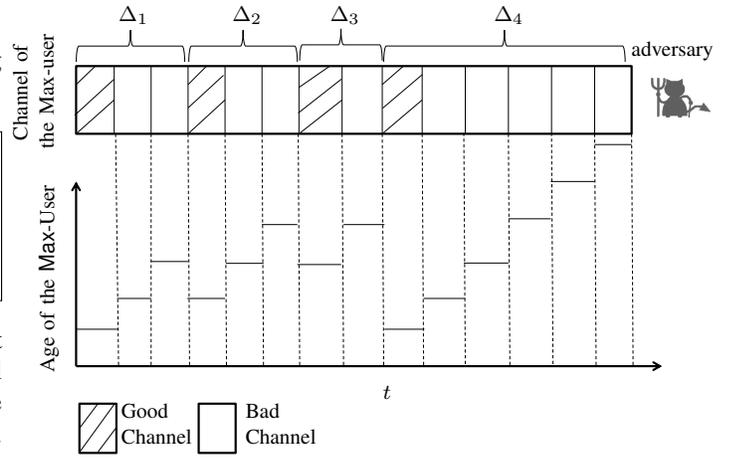
\begin{figure}
\centering
\begin{overpic}[width=0.45\textwidth]{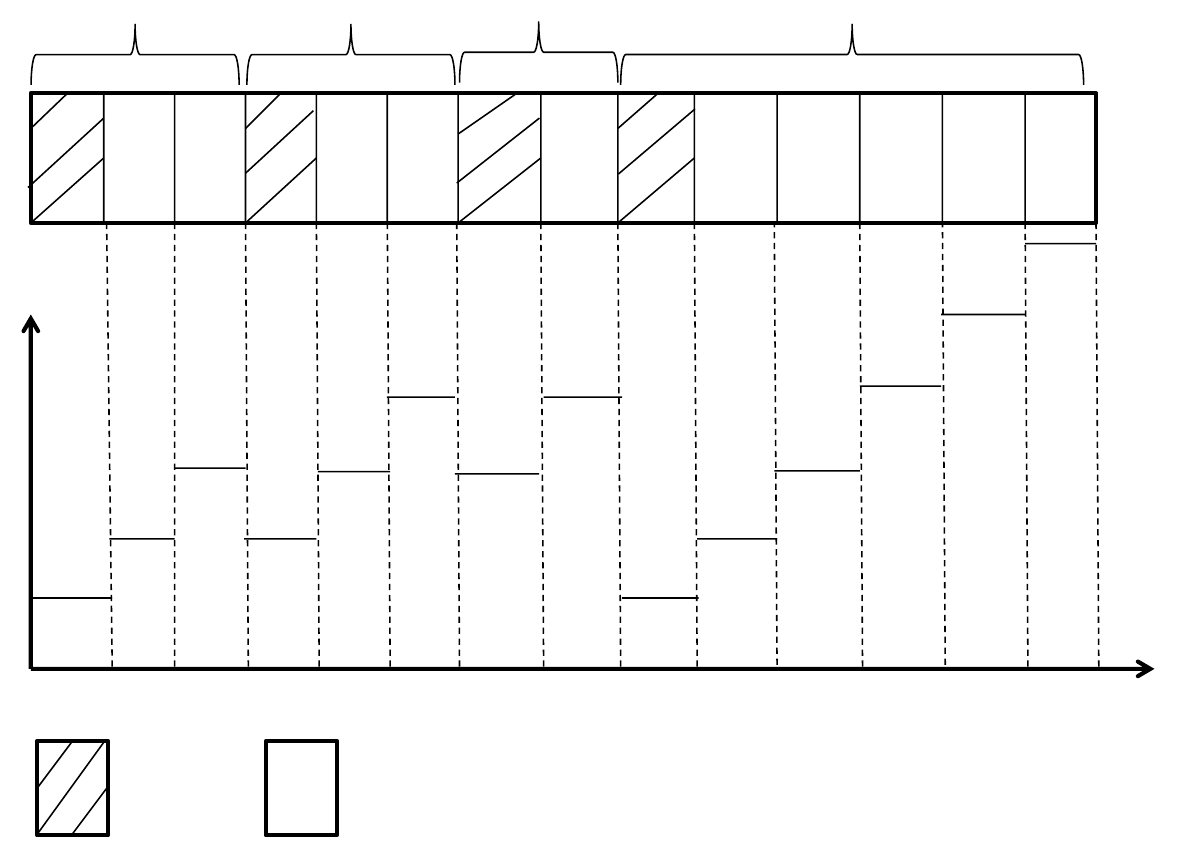}
\end{overpic}
\put(-17,128){\includegraphics[width=0.07\textwidth]{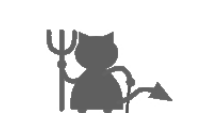}}
\put(-210,168){\footnotesize{$\Delta_1$}}
\put(-16, 155){\footnotesize{adversary}}
\put(-167,168){\footnotesize{$\Delta_2$}}
\put(-130,168){\footnotesize{$\Delta_3$}}
\put(-68,168){\footnotesize{$\Delta_4$}}
\put(-209,18){\footnotesize{Good}}
\put(-209,8){\footnotesize{Channel}}
\put(-162,18){\footnotesize{Bad}}
\put(-110,25){\footnotesize{$t$}}
\put(-162,8){\footnotesize{Channel}}
\put(-240,35){\footnotesize{\rotatebox{90}{Age of the \textsf{Max}-User}}}
\put(-250,123){\footnotesize{\rotatebox{90}{Channel of}}}
\put(-240,120){\footnotesize{\rotatebox{90}{the Max-user}}}
\caption{\small {Illustrating Super-intervals under the \textsf{CMA} policy }} 
\label{intervals_fig}
\end{figure}

For our achievability results, we consider the following online scheduling policy: 
\begin{framed}
\textsf{\underline{Cellular Max-Age (CMA)}}: At every slot, each BS $j$  schedules a downlink packet transmission to the user with the highest age among all users in BS $j$'s cell at that slot (ties are broken in an arbitrary but fixed order).
\end{framed} 
Clearly, the \textsf{CMA} policy is decentralized as the schedulers at each BS need to know the state (AoI) of the users in their local cells only. In the following, we upper bound the competitive ratios of the \textsf{CMA} policy for the Average AoI objective (Eqn.\ \eqref{avg_cost_fn}) and the Peak AoI objective (Eqn.\ \eqref{max_cost_fn}). Surprisingly, it turns out that the bounds are independent of the total number of Base Stations $M$.
 \edit{In the context of the \textsf{CMA} policy}, we now state a few definitions which facilitate the achievability proofs.
\paragraph*{\textsf{Max-user}} For any slot $t$, we define the (global) \emph{\textsf{Max}-user} as the user having the highest age among all $N$ users \edit{under the \textsf{CMA} policy} (ties are broken similarly as in the \textsf{CMA} policy). Clearly, the identity of the \textsf{Max}-user changes with time. Observe that, by definition, the \textsf{CMA} policy continues to schedule the current \textsf{Max}-user \emph{irrespective} of its location until it successfully receives a packet. In the following slot, a different user assumes the role of the \textsf{Max}-user, and the process continues.
\paragraph*{\textsf{Super-interval}} \edit{Under the \textsf{CMA} policy,} the time interval between two consecutive successful transmissions to the current \textsf{Max}-user is called a \emph{super-interval}. Throughout a super-interval, the identity of the \textsf{Max}-user remains invariant. The \textsf{Max}-user corresponding to the $i$\textsuperscript{th} super-interval is denoted by $M_i$. 
Note that the super-intervals are contiguous and disjoint. Let $T_i$ be the index of the time slot at which the $i$\textsuperscript{th} super-interval ends. Thus, $\Delta_i \equiv T_i- T_{i-1}$ denotes the length of the $i$\textsuperscript{th} \emph{super-interval}. See Figure \ref{intervals_fig} for a schematic. Note that there could be more than one successful transmission within a super-interval to users other than the \textsf{Max}-user (by different Base Stations). For notational consistency, we define $T_j \equiv 0, \textrm{and } \Delta_j \equiv 0, \forall j\leq 0.$
The above definitions and observations lead to the following key result:

\begin{framed}
\begin{proposition}\label{max_upper_bound}
At the $k$\textsuperscript{th} slot of the $i$\textsuperscript{th} super-interval, the age of the \textsf{Max}-user $M_i$ under the $\textsf{CMA}$ policy is upper bounded by
    $k+ \sum_{j=1}^{N-1} \Delta_{i-j}.$ 
\end{proposition}
\end{framed}
\begin{IEEEproof}
We claim that the \textsf{Max}-user $M_i$  corresponding to the $i$\textsuperscript{th} super-interval must have had a successful transmission within the last $N-1$ super-intervals. If not, since there are a total of $N$ users, by the pigeonhole principle, some other user $j\neq M_i$ must become the \textsf{Max}-user at least twice in the previous $N$ super-intervals. However, this cannot be true as then, the $j$\textsuperscript{th} user would have had less age than $M_i$ (which was not scheduled at all in the last $N-1$ super-intervals) when the $j$\textsuperscript{th} user became the \textsf{Max}-user for the second time. Thus, at the beginning of the $i$\textsuperscript{th} super-interval, the age of the new \textsf{Max}-user is upper bounded by $\sum_{j=1}^{N-1} \Delta_{i-j}$. The proposition follows from this fact.
\end{IEEEproof}
The previous proposition leads to the following achievability result for the \textsf{CMA} policy: 
\begin{framed}
\begin{theorem}[Upper bounds] \label{avg_comp_ratio_ub}
	The competitive ratios of the \textsf{CMA} policy for the Average AoI and Peak AoI metrics can be upper bounded as follows: \[\eta^{\textsf{CMA}}_{\textrm{avg}} \leq 2N^2, \textrm{ and}~~\eta^{\textsf{CMA}}_{\textrm{peak}} \leq 2N.\]  
\end{theorem}	
\end{framed}
\paragraph*{Discussion on the proof technique}
We prove the bounds for the Average AoI and the  Peak AoI metrics separately. \edit{The achievability proofs proceed by upper bounding the cost incurred by the \textsf{CMA} policy and lower bounding the cost incurred by the \textsf{OPT} policy in each super-interval generated by the \textsf{CMA} policy. Note that the quantities, such as the super-intervals $\{\Delta_i\}_{i\geq 1}$, the number of super-intervals $K$, and the \textsf{Max-User} $M_i$ appearing in the proofs below are defined with reference to the \textsf{CMA} policy only. See Figure \ref{intervals_fig} for an illustration. Hence, these quantities are determined exclusively by the input channel state and user location sequence $\bm{\sigma}$, which remains the same for all policies, including \textsf{OPT}, while computing the competitive ratio via Eqn.\ \eqref{comp_rat_def}. In other words, we first run the \textsf{CMA} policy on the input sequence $\bm{\sigma},$ determine the super-intervals $\{\Delta_i\}_{i=1}^K$ and the corresponding \textsf{Max-users} $\{M_i\}_{i=1}^K$, and then use these quantities to analyze the performance of \textsf{OPT}. We emphasize that if a user $j$ corresponds to the \textsf{Max-user} in the $i$\textsuperscript{th} super-interval, then the user $j$ \emph{need not} have the highest age among all users at the same slot  under the operation of a different scheduling policy (\emph{e.g.,} \textsf{OPT}).
}
\begin{IEEEproof}
\paragraph{Average-AoI}
Note that, at any slot of the $i$\textsuperscript{th} super-interval, the age of every user is upper bounded by that of the \textsf{Max}-user $M_i$. 
Hence, using Proposition \ref{max_upper_bound}, the total cost incurred by the \textsf{CMA} policy during the $i$\textsuperscript{th} super-interval may be upper bounded as:
	\begin{eqnarray*}
	C_i^{\textsf{CMA}} &\equiv & \sum_{t \in {i}^\textsuperscript{th} \textrm{super-interval}}\sum_{i=1}^N h_i(t)\nonumber \\
	 &\leq & \sum_{k=1}^{\Delta_i}N\bigg(k+\big(\sum_{j=1}^{N-1} \Delta_{i-j}\big)\bigg)\nonumber \\
	&= & N \bigg( \frac{\Delta_i(\Delta_i+1)}{2} + \sum_{j=1}^{N-1} \Delta_i \Delta_{i-j} \bigg)\nonumber \\
	&\leq & \frac{N}{2}\bigg( N \Delta_i^2 + \Delta_i+ \sum_{j=1}^{N-1} \Delta_{i-j}^2\bigg),
	\end{eqnarray*}
	where in the last step, we have used the AM-GM inequality to obtain $ \Delta_i \Delta_{i-j} \leq \frac{1}{2}\big(\Delta_i^2 + \Delta_{i-j}^2\big), 1\leq j \leq N-1.$\\
	Let $K$ be the total number of super-intervals in the entire time-horizon of length $T$. The total cost incurred  by the \textsf{CMA} policy for the entire time horizon can be upper bounded as: 
	\begin{eqnarray} \label{CMA1}
	\textrm{Cost}^{\textsf{CMA}}(T)&=& \sum_{i=1}^{K}C_i^{\textsf{MA}} \nonumber \\
	&\leq & \frac{N}{2}\sum_{i=1}^{K}\bigg( N \Delta_i^2 + \Delta_i+ \sum_{j=1}^{N-1} \Delta_{i-j}^2\bigg) \nonumber \\
	&\leq & \frac{N}{2}\sum_{i=1}^{K} \bigg(2N \Delta_i^2 + \Delta_i\bigg).
	\end{eqnarray}
On the other hand, observe that the user $M_i$ must experience \textsf{Bad} channels throughout the $i$\textsuperscript{th} super-interval \edit{generated by the \textsf{CMA} policy.} This is true, as otherwise, the user $M_i$ would have successfully received a packet under the \textsf{CMA} policy, \edit{which, by design, always schedules a packet transmission to the \textsf{Max-user}} irrespective of its location. Hence, the cost incurred by the \textsf{OPT} policy during the $i$\textsuperscript{th} super-interval may be lower bounded by the cost of the user $M_i$ as follows: 
	\begin{eqnarray} \label{COPT1}
	C_i^{\textsf{OPT}} 
	&\geq &\underbrace{(N-1)\sum_{k=1}^{\Delta_i}1}_{\textrm{lower-bound to the cost incurred by other users}} +\nonumber\\
	&&\underbrace{\sum_{k=1}^{\Delta_i} (1+k)}_{\textrm{lower-bound to the cost incurred by } M_i} \nonumber\\
	&\geq & \frac{1}{2} \Delta_i^2 + N\Delta_i. 
	\end{eqnarray}
	In the above, the lower-bound to the cost incurred by $M_i$ is obtained using the fact that, \edit{due to successive \textsf{Bad} channels,} its age increases \emph{linearly} under any policy \edit{(including \textsf{OPT})} throughout the $i$\textsuperscript{th} super interval, starting from the minimum age of at least one. 
Finally, the total cost incurred during the entire horizon of length $T$ is obtained by summing up the cost incurred in the constituent super-intervals. 
	Hence, from Eqns.\ \eqref{CMA1} and \eqref{COPT1}, the competitive ratio $\eta^{\textsf{CMA}}$ of the \textsf{CMA} policy may be upper bounded as:
	\begin{eqnarray*}
	\eta^{\textsf{CMA}}_{\textrm{avg}} &=& \frac{\frac{1}{NT}\sum_{i=1}^K C_i^{\textsf{CMA}}}{\frac{1}{NT}\sum_{i=1}^K C_i^{\textsf{OPT}}}	\\
	&\stackrel{(a)}{\leq}& \frac{\frac{N}{2}\sum_{i=1}^{K} \bigg(2N \Delta_i^2 + \Delta_i\bigg)}{\sum_{i=1}^K \big(\frac{1}{2}\Delta_i^2 + N\Delta_i\big)} \\
	& \leq & 2N^2.
	\end{eqnarray*}

\paragraph{Peak AoI}
The proof proceeds essentially in the same way as the Average Age case. Using proposition \ref{max_upper_bound}, the total cost $C_i^{\textsf{CMA}}$ incurred by the \textsf{CMA} policy during the $i$\textsuperscript{th} super-interval may be upper bounded as: 
 \begin{eqnarray*} \label{CMA3}
	C_i^{\textsf{CMA}} &=& \sum_{t \in i \textrm{\textsuperscript{th} super interval}} \max_{i=1}^N h_i(t)\\
	&\leq& \sum_{k=1}^{\Delta_i} \bigg(k+ \sum_{j=1}^{N-1} \Delta_{i-j}\bigg)
	= \frac{1}{2}\big(\Delta_i^2+\Delta_i) + \sum_{j=1}^{N-1}\Delta_i \Delta_{i-j} \nonumber\\
	&\stackrel{(a)}{\leq}& \frac{1}{2}\big(\Delta_i^2+\Delta_i) + \frac{1}{2}\sum_{j=1}^{N-1}\big(\Delta_i^2 + \Delta_{i-j}^2\big) \label{am_gm}\nonumber \\
	&=& \frac{N}{2}\Delta_i^2 + \frac{1}{2}\Delta_i + \frac{1}{2}\sum_{j=1}^{N-1} \Delta^2_{i-j}. 
 \end{eqnarray*}
where in (a), we have used the AM-GM inequality to conclude $ \Delta_i \Delta_{i-j} \leq \frac{1}{2}\big(\Delta_i^2 + \Delta_{i-j}^2\big), 1\leq j \leq N-1.$
	Hence, assuming that there are $K$ super-intervals in the time-horizon of length $T$, the total cost incurred by the \textsf{CMA} policy over the entire time horizon may be upper bounded as: 
	\begin{eqnarray} \label{CMA2}
	\textsf{Cost}^{\textsf{CMA}}(T)&=& \sum_{i=1}^{K}C_i^{\textsf{CMA}} 
	\leq  \frac{1}{2}\sum_{i=1}^{K} \bigg(2N \Delta_i^2 + \Delta_i\bigg).
	\end{eqnarray}
	On the other hand, the cost incurred by the \textsf{OPT} policy during the $i$\textsuperscript{th} super-interval is trivially lower bounded by the cost of the user $M_i$ during the $i$\textsuperscript{th} super-interval. Note that, as in the previous proof, the user $M_i$ experiences successive \textsf{Bad} channels throughout the $i$\textsuperscript{th} super-interval. Hence, 
	\begin{eqnarray} \label{COPT}
	C_i^{\textsf{OPT}}\geq  \sum_{k=1}^{\Delta_i} (1+k)
	=  \frac{1}{2} \Delta_i^2 + \frac{3}{2}\Delta_i. 
	\end{eqnarray}
Finally, the cost of the entire horizon of length $T$ may be obtained by summing up the cost incurred in each super-intervals. 
	Noting that $\Delta_0=0$, using Eqns.\ \eqref{CMA2} and \eqref{COPT}, the competitive ratio $\eta^{\textsf{CMA}}$ of the \textsf{CMA} policy may be upper bounded as:
	\begin{eqnarray*}
	\eta^{\textsf{CMA}}_{\textrm{peak}} = \frac{\frac{1}{T}\sum_{i=1}^K C_i^{\textsf{CMA}}}{\frac{1}{T}\sum_{i=1}^K C_i^{\textsf{OPT}}}	
	\stackrel{(a)}{\leq} \frac{\frac{1}{2}\sum_{i=1}^{K} \bigg(2N \Delta_i^2 + \Delta_i\bigg)}{\sum_{i=1}^K \big(\frac{1}{2}\Delta_i^2 + \frac{3}{2}\Delta_i\big)} 
	\leq 2N.
	\end{eqnarray*}
\end{IEEEproof}

\subsection*{Tightness of the Bounds}
Next, we show that our analysis of the \textsf{CMA} policy is tight and the bounds given in Theorem \ref{avg_comp_ratio_ub} cannot be improved by more than a constant factor.
\begin{framed}
\begin{theorem} \label{tightness_thm}
	In the adversarial setting described above, we have 
	\begin{eqnarray*}
		\eta^{\textsf{CMA}}_{\textrm{avg}} \geq N^2  \textrm{ and } \eta^{\textsf{CMA}}_{\textrm{peak}} \geq 2N-1.
	\end{eqnarray*}
\end{theorem}
\end{framed}
We defer the proof of the above theorem to Appendix \ref{tightness_thm_proof}.
The proof proceeds by constructing a suitable channel state sequence for $N$ static users in a single cell running the \textsf{CMA} policy. 
\subsection{Minimax Lower Bounds for the Competitive Ratios} \label{competitive_ratio_lb}
We now use Yao's minimax principle to obtain lower bounds to the achievable competitive ratio for minimizing the Average AoI and the Peak AoI metrics. These lower bounds are universal and hold for any online policy. For this, we first recall Yao's minimax principle, which is an immediate consequence of Von Neumann's Minimax Theorem in game theory:
\begin{framed} 
\begin{theorem}[Yao's Minimax principle \cite{albers1996competitive}]
	The competitive ratio of the best randomized online algorithm against an oblivious adversary is equal to the competitive ratio of the best deterministic online algorithm under a worst-case input distribution. 
	\end{theorem}
\end{framed}
From the above principle, it is clear that a lower bound to the competitive ratio under any deterministic online algorithm for a given input distribution $\bm p$ yields a lower bound to the competitive ratio in the adversarial setting, \emph{i.e.,} 
\begin{eqnarray}\label{Yao_lb}
\eta \geq \frac{\mathbb{E}_{\bm{\sigma} \sim \bm{p}}(\textrm{Cost of the Best Deterministic Online Policy})}{\mathbb{E}_{\bm \sigma \sim \bm p}\textrm{(Cost of OPT)}}.	
\end{eqnarray}
Towards this end, we construct a suitable channel-state distribution $\bm{p}$ for $N$ static users located at a single BS. Then, we upper bound the expected cost incurred by \textsf{OPT} and lower bound the expected cost incurred by the \textsf{CMA} policy under the distribution $\bm{p}$ to lower bound $\eta$.
\begin{framed}
\begin{theorem}[Minimax Lower Bounds] \label{comp_ratio_lb}
	The competitive ratio of any online policy $\pi \in \Pi$ with $N$ static users located at a single cell is lower bounded as follows:
	\begin{eqnarray*}
	\eta^\pi_{\textrm{avg}} \geq \frac{N}{2}+ \frac{1}{2N}, ~~ 	\eta^\pi_{\textrm{peak}} \geq \Omega(N/\ln(N)).
	\end{eqnarray*}
\end{theorem}	
\end{framed}
\begin{IEEEproof}
As stated above, to apply Yao's minimax principle (Eqn.\ \eqref{Yao_lb}), we need to (a) lower bound the expected cost incurred by the online policy $\pi$, and (b) upper bound the expected cost incurred by the optimal offline policy for some suitably chosen channel state distribution $\bm{p}$. Selecting a channel state distribution $\bm{p}$, which simultaneously yields a tight lower bound and leads to a tractable analysis, is a non-trivial task. Towards this, we consider the following channel state distribution for $N$ static users located at a single cell.
\paragraph*{Distribution of Channel States $\bm{p}$} At every slot $t$, a user is chosen independently and uniformly at random, and assigned a \textsf{Good} channel. The rest of the $N-1$ users are assigned \textsf{Bad} channels. 

The rationale behind the above choice of the channel state distribution will become clear when we evaluate \textsf{OPT}'s expected cost. In general, the optimal offline policy's cost is obtained by solving a Dynamic Program, which is challenging to analyze. However, with the selected channel state distribution $\bm{p}$, only one user's channel is in \textsf{Good} state at any time. Hence, in this case, the \textsf{OPT} policy schedules the user that has \textsf{Good} channel state for that slot. This leads to a tractable analysis of \textsf{OPT}'s expected cost for both the Average AoI and Peak AoI objectives.

\subsection*{Case I: Average AoI}
\subsubsection{Computation of \textsf{OPT}'s cost}
Let the random variable $C_i(T)$ denote the total cost incurred by the $i$\textsuperscript{th}  user up to time $T$, \emph{i.e.,} 
\begin{eqnarray*}
	C_i(T) = \sum_{t=1}^{T} h_i(t).
\end{eqnarray*}
Hence, the limiting time-averaged expected cost incurred by the \textsf{OPT} policy may be expressed as: 
\begin{eqnarray} \label{opt_ub_1}
\bar{\mathcal{C}}(\textsf{OPT}) \equiv	\lim_{T \to \infty} \frac{1}{T} \sum_{i=1}^{N} \mathbb{E}\big(C_i(T)\big) = \sum_{i=1}^{N} \lim_{T \to \infty} \frac{\mathbb{E}(C_i(T))}{T},  
\end{eqnarray}
In the following, we will show that the above limits exist for the channel state distribution $\bm{p}$.
We now use the Renewal Reward Theorem \cite{gallager2012discrete} to evaluate the RHS of Eqn.\ \eqref{opt_ub_1}. As stated before, since only one channel is in \textsf{Good} state at a slot, the optimal policy \textsf{OPT} simply schedules the user having \textsf{Good} channel.  
It can be verified that, under the \textsf{OPT} policy, for each user $i$, the sequence of age random variables $\{h_i(t)\}_{t\geq 1}$ constitute a renewal process. Clearly, the time instants of scheduling the $i$\textsuperscript{th} user constitutes renewal instants. A generic renewal interval of length $\tau$  for the $i$\textsuperscript{th} user consists of two parts - a sequence of \textsf{Good} channels of length $\tau_\textsf{G}$, followed by a sequence of \textsf{Bad} channels of length $\tau_{\textsf{B}}$. Hence, the AoI cost $c_i(\tau)$ incurred by the user $i$ in any generic renewal cycle may be written as the sum of the costs incurred in two parts:  
 \begin{eqnarray*}
c_i(\tau)&=& c_i(\tau_{\textsf{G}})+ c_i(\tau_{\textsf{B}})\\
&=& \sum_{t=1}^{\tau_{\textsf{G}}}1 +  \sum_{t=1}^{\tau_{\textsf{B}}}(1+t)\\
&=& \tau_{\textsf{G}} + \frac{3}{2}\tau_{\textsf{B}} + \frac{1}{2}\tau_{\textsf{B}}^2. 
\end{eqnarray*}
Let $q\equiv \frac{1}{N}$ be the probability that the channel is \textsf{Good} for the $i$\textsuperscript{th} user at any slot. Hence, from our construction, the random variables $\tau_{\textsf{G}}$ and $\tau_{\textsf{B}}$ follows a Geometric distribution with the following p.m.f. 
\begin{eqnarray*}
\mathbb{P}(\tau_{\textsf{G}}=k) &=& q^{k-1} (1-q), ~~ k \geq 1. \\	
\mathbb{P}(\tau_{\textsf{B}}=k) &=& q(1-q)^{k-1}, ~~ k \geq 1. 
\end{eqnarray*}
Thus, the expected cost incurred by the $i$\textsuperscript{th} user at any renewal cycle is given by: 
\begin{eqnarray}\label{cycle_cost}
	\mathbb{E}(c_i(\tau))= \frac{1}{1-q}+ \frac{3}{2q}+ \frac{2-q}{2q^2}= \frac{1}{q^2(1-q)}.
\end{eqnarray}
Moreover, the expected length of any renewal cycle can be computed to be:
\begin{eqnarray}\label{cycle_length}
\mathbb{E}(\tau)= \mathbb{E}(\tau_{\textsf{G}})+ \mathbb{E}(\tau_{\textsf{B}})= \frac{1}{q(1-q)}.	
\end{eqnarray}

Using Renewal Reward Theorem \cite{gallager2012discrete}, we have 
\begin{eqnarray*}
\lim_{T \to \infty} \frac{\mathbb{E}(C_i(T))}{T}= \frac{\mathbb{E}(c_i(\tau))}{\mathbb{E}(\tau)}=\frac{1}{q}=N, ~~~\forall i.	
\end{eqnarray*}
Hence, from Eqn.\ \eqref{opt_ub_1}, we conclude that the limiting time-averaged total expected cost incurred by \textsf{OPT} is given by 
\begin{eqnarray}\label{opt_ub}
	\bar{\mathcal{C}}(\textsf{OPT})= N^2. 
\end{eqnarray}
\subsubsection{Lower bound to the cost of policy $\pi$}
In order to lower bound the expected cost incurred by any online policy $\pi$ under the distribution $\bm{p}$, we appeal to a special case of Theorem \ref{lb}, stated later in Section \ref{avg_stochastic}. Theorem \ref{lb} gives a lower bound to the average cost incurred by any scheduling policy in a stochastic setting when the channels are modeled as binary erasure channels (\textsf{BEC}) with fixed probabilities of success. Since the proof of Theorem \ref{lb} does not require the channels to be mutually independent, it is applicable to the channel state distribution $\bm{p}$ as well. By plugging in $p_i=\frac{1}{N}, ~\forall i$, and $M=1$ in Theorem \ref{lb}, we conclude that under the channel state distribution $\bm{p}$, the time-averaged expected cost for any online scheduling policy $\pi$ is lower bounded as: 
\begin{eqnarray} \label{lb_gnl}
\bar{\mathcal{C}}(\pi) = \limsup_{T\to \infty} \frac{1}{T}\sum_{i=1}^{N}\mathbb{E}(C_i(T)) \geq \frac{N^3+N}{2}.	
\end{eqnarray}
Finally, using Yao's minimax principle in conjunction with Eqns. \eqref{opt_ub} and \eqref{lb_gnl}, we conclude that the competitive ratio $\eta^{\pi}_{\textrm{avg}}$ of any online scheduling policy $\pi$ is lower bounded as: 
\begin{eqnarray}\label{avg_lb}
\eta^{\pi}_{\textrm{avg}} \geq \sup_{T}\frac{C_T(\pi)}{C_T(\textsf{OPT})}	\geq \limsup_{T \to \infty}\frac{C_T(\pi)/T}{C_T(\textsf{OPT})/T} \geq \frac{N}{2} + \frac{1}{2N}. 
\end{eqnarray}
We point out that the lower bound in Eqn.\ \eqref{avg_lb} can be further improved in the case of $N=2$ users using a more careful analysis. The following result shows that the lower bound for average AoI for $N=2$ users may be improved to $1.5$ from $1.25$ as given by Theorem \ref{comp_ratio_lb}. 
\begin{framed}
\begin{proposition} \label{improved_LB}
For the case of $N=2$ users, we have the following improved bound: $\eta^{\pi}_{\textrm{avg}} \geq 1.5.$
\end{proposition}
\end{framed}
Please refer to Appendix \ref{improved_LB_proof} for the proof. \editr{Proposition \ref{improved_LB} suggests that the lower bound on the competitive ratio for the average age metric is likely to be loose and may be improved upon further using a more refined analysis.}  \\
Next, we consider the Peak-AoI objective \eqref{max_cost_fn} and derive a minimax lower bound for this cost metric.
\subsection*{Case II- Peak AoI}
\subsubsection{Upper bound to OPT's cost} 
We use the same channel state distribution $\bm{p}$ as before. Recall that, under the distribution $\bm{p}$, at any slot $t$:
\begin{eqnarray*}
\mathbb{P}(\textrm{user } i\textrm{'s channel is \textsf{Good} at slot~} t)&=&1/N,\\
\mathbb{P}(\textrm{user } i\textrm{'s channel is \textsf{Bad} at slot~} t)&=&1-1/N,
\end{eqnarray*}
independent of everything else.
The \textsf{OPT} policy, with non-causal channel state information, schedules the user having a \textsf{Good} channel at every slot. Thus, the limiting distribution of the age of any user is Geometric ($\frac{1}{N}$), \emph{i.e.,}
\begin{eqnarray*}
\lim_{t \to \infty} \mathbb{P}\big(h_i(t)=k\big) = \frac{1}{N}\bigg(1-\frac{1}{N}\bigg)^{k-1}, ~~ k \geq 1, \forall i.	
\end{eqnarray*}
Hence, for upper bounding the time-averaged cost of \textsf{OPT} using Ces\`{a}ro's summation formula, we are required  to  upper-bound the expected value of maximum of $N$ \emph{dependent} and identically Geometrically distributed random variables.
The MGF of the Geometric distribution $G$ is given by: 
\begin{eqnarray*}
\mathbb{E}(\exp(\lambda G)) =\begin{cases}
 	\frac{e^\lambda/N}{1-e^\lambda(1-1/N)},~~\textrm{ if } \lambda < -\log(1-1/N)\\
 	\infty ~~ \textrm{o.w.}
 \end{cases}
\end{eqnarray*}
Let the random variable \ $H_{\max}$ denote limiting peak-age of the users. 
 For any $-\log(1-1/N)>\lambda >0,$ we have:
\begin{eqnarray*}
	&&\exp\big(\lambda \mathbb{E}(H_{\max})\big)\\
	&\stackrel{(a)}{\leq}&  \mathbb{E}(\exp(\lambda H_{\max})) 
	\leq \sum_{i=1}^N \mathbb{E}(\exp(\lambda G_i)) 
	\leq \frac{e^\lambda}{1- e^\lambda (1-\frac{1}{N})},
\end{eqnarray*}
where the inequality (a) follows from Jensen's inequality.
Taking natural logarithm of both sides, we get
\begin{eqnarray}\label{H_max_bd}
 \mathbb{E}(H_{\max}) \leq 1 - \frac{1}{\lambda}\log\big(1-e^\lambda (1-1/N)\big).	
\end{eqnarray}
Now, let us choose $\lambda= \frac{\alpha}{N},$ for some fixed $\alpha$ $(0< \alpha <1)$ that will be fixed later. First, we verify that, with this choice for $\lambda$, we always have $\lambda < -\log(1-\frac{1}{N})$. Using the fact that $e^{-x} \geq 1-x, \forall x,$ we have 
\begin{eqnarray} \label{cvx_bd}
e^x \leq \frac{1}{1-x}, ~~\forall x<1.
\end{eqnarray}
As a result,
\begin{eqnarray*}
e^\lambda\equiv e^{\frac{\alpha}{N}} \leq \frac{1}{1-\frac{\alpha}{N}} < \frac{1}{1-\frac{1}{N}}; \textrm{ i.e., } \lambda < -\log(1-\frac{1}{N}). 	
\end{eqnarray*}
Next, for upper bounding the RHS of Eqn.\  \eqref{H_max_bd}, we start with the simple analytical fact that for any $0<\alpha <1,$
\begin{eqnarray}\label{analysis1}
	\inf_{0<x<1} \frac{1-(1-x)e^{\alpha x}}{x} = 1-\alpha.
\end{eqnarray}
This result can be verified by using Eqn.\ \eqref{cvx_bd} to conclude that for any $0<x<1,$ we have
\begin{eqnarray*}
	\frac{1-(1-x)e^{\alpha x}}{x} \geq \frac{1}{x}\big(1- \frac{1-x}{1-\alpha x}\big)= \frac{1-\alpha}{1-\alpha x} \geq 1-\alpha,
\end{eqnarray*}
where the infimum is achieved when $x\to 0^+.$ Substituting $x =\frac{1}{N}$ in the inequality \eqref{analysis1}, we have 
\begin{eqnarray*}
1-e^{\alpha/N}(1-1/N)\geq \frac{1-\alpha}{N}.	
\end{eqnarray*}
Hence, using Eqn.\ \eqref{H_max_bd}, we have the following upper bound to the expected Max-age under \textsf{OPT}:
\begin{eqnarray*}
\mathbb{E}(H_{\max})\leq 1+ \frac{N}{\alpha}\ln \frac{N}{1-\alpha},	\textrm{ for some } 0< \alpha <1.
\end{eqnarray*}
Setting $\alpha = 1-\frac{1}{\ln N}$ yields the following asymptotic bound: 
\begin{eqnarray}\label{lb1}
	\mathbb{E}(H_{\max}) \leq N\ln N + o(N\ln N). 
\end{eqnarray}

\subsubsection{Lower Bound to the expected cost of any online policy $\pi$}
To establish a lower bound to the expected cost of any online policy $\pi$, we use Theorem \ref{opt}, established in Section \ref{max_stochastic} . Theorem \ref{opt} gives the minimum time-averaged peak-AoI cost in the stationary environment when all channels can be modeled as \textsf{BEC}. As in the average AoI case, it can be verified that the theorem continues to hold under the channel state distribution $\bm{p}$. Taking $p_i=\frac{1}{N}, ~\forall i$, and $M=1$ in Theorem \ref{opt}, we conclude that under the distribution $\bm{p}$, the time-averaged expected cost for any online scheduling policy $\pi$ is lower bounded as:
\begin{eqnarray} \label{lb2}
	\liminf_{T \to \infty}\frac{1}{T}\mathbb{E}\textsf{Cost}^{\pi}(T)=\liminf_{T \to \infty} \frac{1}{T}\sum_{t=1}^T \mathbb{E}(\max_i h^{\pi}_i(t)) \geq N^2.
\end{eqnarray}
Combining Eqns.\ \eqref{lb1} and \eqref{lb2} with Eqn.\ \eqref{Yao_lb} and using Ces\`{a}ro's summation formula, we have for any online policy $\pi \in \Pi$: 
\begin{eqnarray*}
\eta^{\pi}_{\textrm{peak}}  
&\geq& \sup_{T >0} \frac{\mathbb{E}\textsf{Cost}^{\pi}(T)}{\mathbb{E}\textsf{Cost}^{\textsf{OPT}}(T)} \\
&\geq& \limsup\limits_{T \to \infty} \frac{\mathbb{E}\textsf{Cost}^{\pi}(T)/T}{\mathbb{E}\textsf{Cost}^{\textsf{OPT}}(T)/T} \\ 
&\geq& \Omega(\frac{N}{\ln N}).
\end{eqnarray*}
\end{IEEEproof}
\paragraph*{Discussion and open problems}\editr{Combining the above results, we see that for the average AoI metric, there is a factor of $O(N)$ gap between the competitive ratio of the \textsf{CMA} policy and the corresponding lower bound. Theorem \ref{tightness_thm} shows that the upper bound to the competitive ratio of the \textsf{CMA} policy, given by Theorem \ref{avg_comp_ratio_ub}, is tight and cannot be improved further (up to a constant factor). Hence, either there exists a  different online policy with a smaller competitive ratio, or the current lower bound for average AoI can be improved with a more involved analysis. Reducing the current optimality gap for the average AoI metric is an interesting open problem. }

\section{AoI Minimization in Stochastic Environments}\label{stochastic}

\begin{table*}
\caption{Summary of the results for the Stochastic Setting} \label{summary1}
\centering
\begin{tabular}{ccccccc}
\toprule
\textbf{Metrics} & \textbf{Cost function} & \textbf{Mobility} &\textbf{Upper Bound} & \textbf{Attaining Policy} &\textbf{Lower Bound}  & \textbf{Approx.} \\
\midrule
Average AoI  & $\limsup_{T \to \infty} \frac{1}{NT} \sum_{t=1}^{T}\bigg(\sum_{i=1}^{N} \mathbb{E}^\pi(h_i(t))\bigg)$        &\textsf{Yes} &$\frac{N}{Mp\big(1-(1-M^{-1})^N\big)}$ & \textsf{MMW} &$\frac{1}{2N g(\bm{\psi})}	\bigg(\sum_{i=1}^{N} \sqrt{\frac{1}{p_i}}\bigg)^2+ \frac{1}{2}$ & $2$   \\
\midrule
Peak AoI & $ \limsup_{T\to \infty} \frac{1}{T}\sum_{t=1}^{T} \mathbb{E}(\max_i h_i(t))$  & \textsf{No}  &$\sum_{i=1}^{N} \frac{1}{p_i}$& \textsf{CMA}  & $\sum_{i=1}^{N} \frac{1}{p_i}$ & $1$\\
\bottomrule
\end{tabular}
\label{stoch-table}
\end{table*} 
In this section, we study the problem of AoI-optimal multi-user scheduling in a cellular wireless network when the channel and user mobility can be modeled as a stationary stochastic process. In the following, we highlight the major differences between the adversarial model in Section \ref{adversarial} and the stochastic model in this section.
Our main results for the stochastic model are summarized in Table \ref{summary1}.

\subsection*{Stochastic System Model}

\paragraph{Channel model} As in the adversarial model, we consider a cellular system where  $N$ users roam in an area having $M$ Base Stations. However, contrary to the adversarial model, the wireless link between the $i$\textsuperscript{th} user and the associated BS is modeled using a stationary binary erasure channel (\textsf{BEC}) with the probability of successful reception of a transmitted packet being $p_i, 0< p_i \leq 1, \forall i$.  Hence, when the associated BS schedules a downlink packet transmission to the $i$\textsuperscript{th} user, the packet is either successfully received with probability $p_i$ (if the channel is in \textsf{Good} state) or is lost otherwise (if the channel is in \textsf{Bad} state). Due to the power control mechanisms employed by Base Stations, the success probabilities (\emph{i.e.,} the parameter $p_i$'s) may vary among the users \cite{powercontrol, tse2005fundamentals}. The channels are i.i.d. with respect to the time but need not be independent across the users. 

\paragraph{Mobility model} Contrary to the adversarial setting, where we allow arbitrary mobility patterns, in the stochastic setting, we model the user mobility using a stationary ergodic process. Formally, let the random variable $C_i(t) \in \{1,2,\ldots, M\}$ denote the index of the cell to which the $i$\textsuperscript{th} user is associated with at time $t$. We assume that the stochastic process $\{C_i(t)\}_{t \geq 1}$ is a stationary ergodic process such that $\mathbb{P}(C_i(t)=j)= \psi_{ij}, \forall i \in [N],j \in [M], t\geq 1.$ The probability measure $\bm{\psi}$ denotes the time-invariant occupancy distribution of the cells by the users. The mobility of the users may be correlated or independent of each other. Many different stochastic mobility models proposed in the literature fall under the above general scheme, including the i.i.d. mobility model, random walk model, and the random waypoint model \cite{ge2016user, akyildiz2000new, johnson1996dynamic, bai2004survey}. 

\paragraph{Packet arrival model and Policy Space} 
We use the same saturated traffic model for packet arrivals as the adversarial setup. The policy space $\Pi$ is also identical to the adversarial framework. In particular, we study online scheduling policies that know the channel parameters $\{p_i\}_{i=1}^N$ but do not know the current or future realizations of random channel states. 


\paragraph{Performance metrics}
 For a given policy $\pi \in \Pi,$ its long-term Average AoI metric is defined as:
\begin{eqnarray} \label{avg_objective}
\textsf{AoI}_{\textrm{avg}}^\pi= \limsup_{T \to \infty} \frac{1}{T} \sum_{t=1}^{T}\frac{1}{N}\bigg(\sum_{i=1}^{N} \mathbb{E}^\pi(h_i(t))\bigg).
\end{eqnarray}
Similarly, the long-term Peak AoI metric achieved by a policy $\pi \in \Pi$ is defined as:
\begin{eqnarray}\label{max_objective}
\textsf{AoI}_{\textrm{peak}}^\pi=\limsup_{T\to \infty} \frac{1}{T}\sum_{t=1}^{T} \mathbb{E}(\max_i h_i(t)).
\end{eqnarray}
 Our goal is to design scheduling policies that minimize the long-term expected Average AoI and the long-term expected Peak AoI.
\edit{\paragraph*{Discussion} Our stochastic model is simplistic in the sense that it does not account for time-varying path losses. A more realistic model would take into account the time-varying nature of successful packet reception probabilities as a function of the users' current distances from the associated BS. However, such a model is challenging to analyze and not expected to shed much additional insight into the design of good scheduling policies. Our choice of the above simple model is motivated by the following two reasons:
\begin{enumerate}
\item In practice, dynamic power control mechanisms employed in cellular networks tightly regulate the transmission power from the BS to each user so that the received SNR levels remain almost constant \cite{tse2005fundamentals, chiang2008power, zappone2015energy}. This keeps the probability of successful packet receptions for each user roughly invariant irrespective of their locations and thus, mitigates the cell-edge effects. 

\item Although our stochastic model is simple, it sheds valuable insights into the design of good scheduling policies even in non-stationary environments. In Section \ref{simulation}, we show that the effect of time-varying channel parameters can be heuristically incorporated into the proposed policies, which lead to improved performance. 
\end{enumerate} }

\edit{\paragraph*{Proof Techniques}
The problem of finding an optimal policy for minimizing the AoI in the stochastic setting (Eqns.\ \eqref{avg_objective} and \eqref{max_objective}) reduces to solving an infinite state, unbounded cost Markov Decision Process (MDP) problem with an average cost objective. Such problems are notoriously difficult to tackle, and at present, there is no comprehensive theory (see Section 4.6 of \cite{bertsekas1995dynamic}). Moreover, standard numerical approximation schemes for infinite-state MDPs do not typically provide explicit performance guarantees \cite{ferns2012metrics}.
 As a result, solving the average cost MDP problems with infinite states require problem-specific techniques.

 For the problem of minimizing the peak AoI in Section \ref{max_stochastic}, we directly establish the optimality of a greedy index policy in Theorem \ref{opt} by making use of a clever guess for the differential value function in Bellman's equation. However, a similar direct method could not be found for the problem of minimizing the average AoI metric. Hence, for this problem, we resort to an indirect approach. In Theorem \ref{lb}, we establish a lower bound to the average AoI incurred by \emph{any} admissible scheduling policy. Following this result, in Theorem \ref{achievability_thm}, we establish that the proposed \textsf{MMW} policy achieves an average AoI, which is at most twice as large as the above lower bound under certain modelling assumptions. 
 
}

\subsection{Minimizing the Average AoI in the Stochastic Setting}\label{avg_stochastic}
\subsubsection{Lower bound}
 The following theorem gives a universal lower bound to the average AoI metric attained by any admissible policy. 
\begin{framed} 
\begin{theorem}[Converse] \label{lb}
In the stochastic setup, the optimal average AoI is lower bounded as:
 \begin{eqnarray} \label{lb_expr}
 \textsf{AoI}_{\textrm{avg}}^* \geq 	\frac{1}{2N g(\bm{\psi})}	\bigg(\sum_{i=1}^{N} \sqrt{\frac{1}{p_i}}\bigg)^2+ \frac{1}{2},
 \end{eqnarray}
where the function $g(\bm{\psi})$ denotes the expected number of cells having at least one user under the stationary occupancy distribution $\bm{\psi}$. 
\end{theorem}
\end{framed}
Since $g(\bm \psi) \leq  \min \{M, N\},$ we also have the following weakening of the above bound, which is agnostic of the user mobility statistics:
 \begin{eqnarray*}
 \textsf{AoI}_{\textrm{avg}}^* \geq 	\frac{1}{2N \min \{M, N\}}	\bigg(\sum_{i=1}^N \sqrt{\frac{1}{p_i}}\bigg)^2 + \frac{1}{2}.
 \end{eqnarray*} 

\begin{IEEEproof}
We use a sample-path-based argument to obtain an almost sure lower bound to the average AoI. We then use Fatou's lemma \cite{williams1991probability} to convert the almost sure bound to a bound in expectation. 

Consider a sample path under the action of any arbitrary admissible scheduling policy $\pi \in \Pi$. Let the random variable $N_i(T)$ denote the number of packets received by the $i$\textsuperscript{th} user up to time $T$. Also, let the random variable $T_{ij}$ denote the time interval between receiving the $(j-1)$\textsuperscript{th} packet and the $j$\textsuperscript{th} packet, and the random variable $D_i$ denote the time interval between receiving the last (\emph{i.e.,} $N_i(T)$\textsuperscript{th}) packet and the time-horizon $T$ for the $i$\textsuperscript{th} user. Hence, we can write
\begin{eqnarray} \label{sum_val}
	T= \sum_{j=1}^{N_i(T)} T_{ij} + D_i.
\end{eqnarray}

Since the AoI of any user increases by one at each slot until a new packet is received and then it drops to one again (Figure \ref{AoI_fig}), the average AoI up to time $T$ may be lower bounded as:
\begin{eqnarray} \label{AoI_lb_der}
	\overline{\textsf{AoI}_{T}}&\equiv&\frac{1}{NT}\sum_{i=1}^{N} \sum_{t=1}^{T} h_i(t) \nonumber \\
	 &= & \frac{1}{NT}\sum_{i=1}^{N}\bigg(\sum_{j=1}^{N_i(T)} \frac{1}{2}T_{ij}(T_{ij}+1)+ \frac{1}{2}D_i(D_i+1)\bigg) \nonumber \\
	&\stackrel{(a)}{=}&\frac{1}{2NT}\sum_{i=1}^{N}\bigg(N_i(T) \big(\frac{1}{N_i(T)} \sum_{j=1}^{N_i(T)}T_{ij}^2 \big)+D_i^2\bigg)+ \frac{1}{2}\nonumber\\
	&\stackrel{(b)}{\geq} & \frac{1}{2NT}\sum_{i=1}^{N}\bigg( N_i(T)\bar{T_i}^2+D_i^2\bigg)+ \frac{1}{2},
\end{eqnarray}
where in Eqn.\ (a) we have used \eqref{sum_val}, and in Eqn.\ (b) 
we have defined $\bar{T}_i \equiv \frac{1}{N_i(T)} \sum_{j=1}^{N_i(T)} T_{ij}$ and used Jensen's inequality. Rearranging Eqn.\ \eqref{sum_val}, we can express the random variable $\bar{T}_i$ as: 
\begin{eqnarray*}
	\bar{T}_i= \frac{T-D_i}{N_i(T)}.
\end{eqnarray*}

With this substitution, the term within the bracket in Equation \eqref{AoI_lb_der} simplifies to 
\begin{eqnarray} \label{AoI_lb_der2}
	 N_i(T)\bar{T}_i^2+ D_i^2 = \frac{(T-D_i)^2}{N_i(T)} + D_i^2 \geq \frac{T^2}{N_i(T)+1},
\end{eqnarray}
where the last inequality is obtained by minimizing the middle expression by viewing it as a quadratic in the variable $D_i$. \\
Hence, from Eqns.\ \eqref{AoI_lb_der} and \eqref{AoI_lb_der2}, we obtain the following lower bound to the average AoI under the action of any admissible scheduling policy: 
\begin{eqnarray} \label{val}
	\overline{\textsf{AoI}_T} \geq \frac{T}{2N} \sum_{i=1}^{N} \frac{1}{N_i(T)+1} + \frac{1}{2}.
\end{eqnarray}
Next, we incorporate the scheduling constraints to control the RHS of inequality \eqref{val}. Let the random variable $A_i(T)$ denote the total number of transmission attempts made to the $i$\textsuperscript{th} user by all Base Stations up to time $T$. Also, let the random variable $g_j(T)$ denote the fraction of time that $\textrm{BS}_j$ contained at least one user in its coverage area. Since a BS can attempt a downlink transmission only when there is at least one user in its coverage area, the total number of transmission attempts to all users by the Base Stations can be upper bounded by the following global balance condition:
%
%
\begin{eqnarray} \label{attmpt_constr}
\sum_{i=1}^{N} A_i(T) \leq T \sum_{j=1}^{M} g_j(T)\equiv T g(T), 	
\end{eqnarray}
where $g(T) \equiv \sum_j g_j(T)$. Using inequality \eqref{attmpt_constr}, we can lower bound the RHS of inequality \eqref{val} as:
 \begin{eqnarray}
\overline{\textsf{AoI}_T} \geq \frac{1}{2Ng(T)} \big(\sum_{i=1}^{N} A_i(T)\big)\big(\sum_{i=1}^{N} \frac{1}{N_i(T)+1}\big) + \frac{1}{2}.
\end{eqnarray}
An application of the Cauchy-Schwartz inequality on the RHS of the above inequality yields:
\begin{eqnarray} \label{AoI_lb_2}
\overline{\textsf{AoI}_T} \geq\frac{1}{2N g(T)}\bigg(\sum_{i=1}^{N} \sqrt{\frac{A_i(T)}{N_i(T)+1}}\bigg)^2 + \frac{1}{2}.
\end{eqnarray}
 Note that, the $i$\textsuperscript{th} user successfully received  $N_i(T)$ packets out of a total of $A_i(T)$ packet transmission attempts made by the Base Stations over the wireless erasure channels with success probability $p_i$. Without any loss of generality, we may fix our attention on those scheduling policies only for which $\lim_{T \to \infty} A_i(T)= \infty, \forall i$ almost surely. Otherwise, at least one of the users will receive only a finite number of packets, resulting in infinite average AoI. Hence, using the Strong law of large numbers \cite{williams1991probability}, we obtain:  
 \begin{eqnarray} \label{SLLN2}
 \lim_{T \to \infty} \frac{N_i(T)}{A_i(T)} = p_i, ~~~\forall i \hspace{5pt} \textrm{w.p.} ~ 1. 	
 \end{eqnarray}
Moreover, using the ergodicity property of the user mobility, we conclude that almost surely: 
\begin{eqnarray*}
\lim_{T \to \infty} g_j(T) = \mathbb{P}_{\bm{\psi}}\big( \textrm{BS}_j \textrm{ contains at least one user}\big),	
\end{eqnarray*}
where we recall that $\bm{\psi}$ denotes the stationary cell occupancy distribution. Thus, we have almost surely 
\begin{eqnarray} \label{g_lim}
\lim_{T \to \infty} g(T) &=& \lim_{T\to \infty} \sum_j g_j(T)\nonumber\\
&=&  \sum_{j=1}^{M} \mathbb{P}_{\bm \psi} \big( \textrm{BS}_j \textrm{ contains at least one user}\big) \nonumber \\
&\equiv& g(\bm{\psi}), 	
\end{eqnarray}
where the function $g(\bm{\psi})$ denotes the expected number of non-empty cells under the stationary occupancy distribution $\bm \psi$. Hence, combining equations \eqref{SLLN2} and \eqref{g_lim} together with the lower bound in equation \eqref{AoI_lb_2}, we have almost surely:
\begin{eqnarray} \label{AoI_LB2}
\liminf_{T \to \infty} \overline{\textsf{AoI}_T} \geq 
\frac{1}{2N g(\bm{\psi})}	\bigg(\sum_i \sqrt{\frac{1}{p_i}}\bigg)^2 + \frac{1}{2}.
\end{eqnarray}
Finally, 
\begin{eqnarray*}
\textsf{AoI}_{\textrm{avg}}^* &\geq& \liminf_{T \to \infty} \mathbb{E}(\textsf{AoI}_T) \\
&\stackrel{(a)}{\geq}& \mathbb{E}(\liminf_{T \to \infty} \textsf{AoI}_T) \\
&\geq& 	\frac{1}{2N g(\bm{\psi})}	\bigg(\sum_i \sqrt{\frac{1}{p_i}}\bigg)^2 + \frac{1}{2},
\end{eqnarray*} 
where the inequality (a) follows from Fatou's lemma. This concludes the proof of Theorem \ref{lb}. Note that the proof continues to hold even when the mobility of the users is not independent across the users. 
\end{IEEEproof}

\paragraph*{Discussion} 
Theorem \ref{lb} suggests that the user mobility statistics affects the lower bound only through the stationary cell-occupancy distribution $\bm{\psi}$. Hence, given the stationary distribution $\bm \psi$, the lower bound \eqref{lb_expr} is agnostic of the specifics of the mobility model. A similar result was obtained earlier in connection with the capacity region of wireless networks (see \cite{neely2003dynamic}, Corollary $5$, p. 88). 
\edit{These observations can be intuitively understood as follows. Note that, we are interested in the asymptotic AoI of the users averaged over an arbitrarily long time-horizon (\emph{viz.} Eqn.\ \eqref{avg_objective}). For many ``regular" stochastic processes, \emph{e.g.,} positive recurrent Markov Chains or general Ergodic processes,  it is well-known that the long-term behaviour of the process is entirely determined by its limiting steady-state distribution (see, \emph{e.g.,} Theorem 7.2.1 of \cite{durrett2010probability}). In practice, we are interested in scheduling policies that are ``well-behaved" in the above sense. Hence, it is not surprising that the lower bound \eqref{lb_expr}, which is used to upper bound the approximation ratio of the scheduling policies, depends only on the induced steady-state distribution and not on the details of the transition probabilities of the specific mobility model.} The appearance of the quantity $g(\bm \psi)$ in the lower bound should not be surprising either as it denotes the \emph{typical} number of non-empty cells at a time in the long run. Since a Base Station can transmit a packet only if at least one user is present in its coverage area, the quantity $g(\bm \psi)$, in some sense, represents the multi-user diversity of the system.

\subsubsection*{Closed-form expression for $g(\bm \psi)$}
To get a sense of the bound \eqref{lb_expr}, we now derive a closed-form expression for $g(\bm \psi)$ under certain assumptions. This expression will be used later in Theorem \ref{achievability_thm} to establish the $2$-optimality guarantee of the \textsf{MMW} policy discussed in the following section. 

\begin{figure}
 \centering
 \includegraphics[draft=false, scale=0.18]{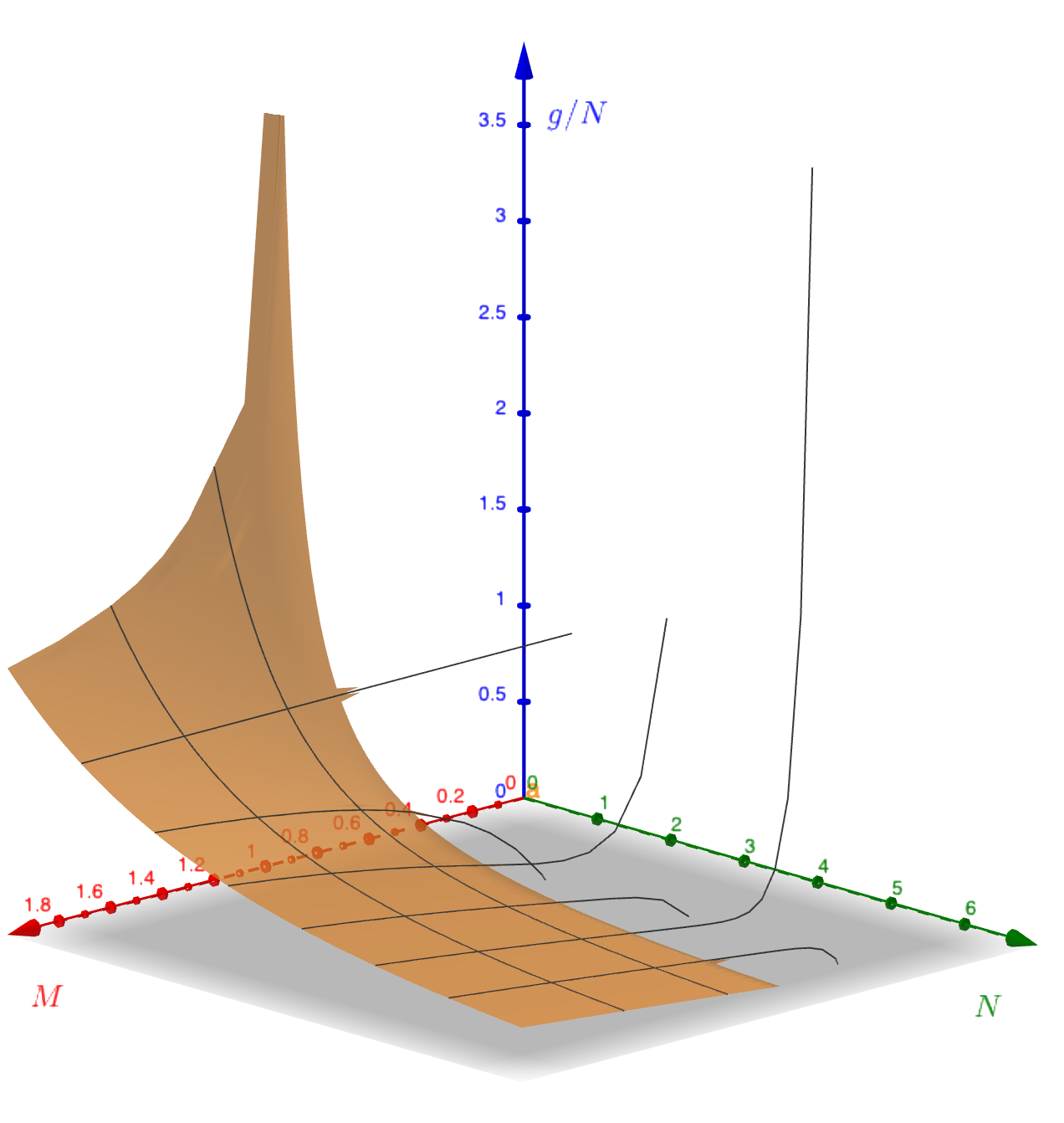}
 \caption{Plot of $g(\psi)/N$ as a function of $M$ and $N$. Note that the average AoI is inversely proportional to $g(\psi)/N$ in the case when all users are identical (\emph{i.e.,} $p_i=p, \forall i$).}	
 \label{g-psi-plot}
 \end{figure}

Using the linearity of expectation, we have
\begin{eqnarray}\label{g_psi_eq}
	g(\bm{\psi})&=& \mathbb{E}_{\bm \psi} \sum_{j=1}^{M} \mathds{1}(\textrm{BS}_j \textrm{ contains at least one user}\big)\nonumber \\ 
	&=&\sum_{j=1}^{M} \mathbb{P}_{\bm \psi} \big( \textrm{BS}_j \textrm{ contains at least one user}\big). 
\end{eqnarray}
Since the cells are disjoint, we readily conclude from Eqn.\ \eqref{g_psi_eq} that $g(\bm{\psi}) \leq \min\{M, N\}$. Recall that $\psi_{ij}$ denotes the probability that the $i$\textsuperscript{th} user is in $\textrm{BS}_j$. If the mobility of the users is independent of each other, the expected number of non-empty cells $g(\bm \psi)$ in Eqn.\ \eqref{g_psi_eq} simplifies to:
\begin{eqnarray} \label{g_eqn3}
g(\bm{\psi})= \sum_{j=1}^{M} \big(1-\prod_{i=1}^N(1-\psi_{ij})\big).	
\end{eqnarray}
%
%
%
%
We now evaluate the above expression for the 
 case when the limiting occupancy distribution of each user is uniform across all cells, \emph{i.e.,} $\psi_{ij}=\frac{1}{M}, \forall i,j $. The uniform stationary distribution arises, for example, when the user mobility can be modelled as a random walk on a regular graph \cite{lovasz1993random}. In this case, Eqn.\ \eqref{g_eqn3} simplifies to
 \begin{eqnarray} \label{g_spl}
g(\bm{\psi^{\textsf{unif}}}) = M \bigg(1-\big(1-\frac{1}{M}\big)^N\bigg).  	
\end{eqnarray}
For $M=1$, we have $g(\bm{\psi})=1$. For $M \geq 2$, we have the following bounds: 
\begin{eqnarray} \label{g_ineq1}
 e^{-\frac{\beta}{M}} \stackrel{(a)}{\leq} (1-\frac{1}{M}) \stackrel{(b)}{\leq} e^{-\frac{1}{M}},  
\end{eqnarray}
where $\beta \equiv \log(4) \leq 1.387.$ 
The inequality (b) is standard. To prove the inequality (a), consider the concave function 
\[f(x) = 1-x-e^{-\beta x}, 0\leq x \leq \frac{1}{2},\]
 for some $\beta > 0$. Since a concave function of a real variable defined on an interval attains its minima at one of the end points of the closed interval, and since $f(0)=0$, we have $f(x) \geq 0, \forall x \in [0, \frac{1}{2}],$ if $f(1/2) \geq 0$, i.e., $ e^{\beta /2} \geq 2$, i.e., $\beta \geq \ln(4)$. Thus, the inequality (a) holds for $M\geq 2$ with $\beta = \ln(4)$. The inequality \eqref{g_ineq1} directly leads to the following bounds for $M \geq 2$:
 \begin{eqnarray} \label{g_unif}
 M\bigg(1-e^{-\frac{N}{M}}\bigg)\leq g(\bm{\psi^{\textsf{unif}}}) \leq M\bigg(1-e^{-1.387 \frac{N}{M}}\bigg).	
 \end{eqnarray}
 \edit{Please refer to Figure \ref{g-psi-plot} for the plot of the normalized $g(\cdot)$ function.}

 \subsubsection{Achievability}
 
 We now propose an online scheduling policy, called ``Multi-cell Max-Weight" or $\pi^{\textsf{MMW}},$ that approximately minimizes the long-term average AoI. 
 \begin{framed}
\emph{The policy $\pi^{\textsf{MMW}}$}:  At each slot $t$, each BS transmits a packet to a user in its cell that has the highest index among all other users in its cell. The index $I_i(t)$ of the $i$\textsuperscript{th} user at time $t$ is defined as $I_i(t) \equiv p_ih_i^2(t), \forall i \in [N].$
\end{framed}
  The policy $\pi^{\textsf{MMW}}$ is a multi-cell generalization of the $4$-approximation single cell scheduling policy $\textsf{MW}$ proposed in \cite{kadota2018scheduling}. In Theorem \ref{achievability_thm}, we show that $\pi^{\textsf{MMW}}$ is a $2$-approximation policy for statistically identical users with i.i.d. uniform mobility.
  The policy $\pi^{\textsf{MMW}}$ is also a $2$-approximation policy for static users located in a single cell (possibly with varying transmission success probabilities). Hence, our result strictly improves upon the $4$-approximation guarantee of the \textsf{MW} policy for static users \cite{kadota2018scheduling}. Our result also complements
  Theorem 5 of \cite{kadota2018scheduling}, where the authors showed that the \textsf{MW} policy is exactly optimal for statistically identical static users located in a common cell.  
 

\begin{framed}
\begin{theorem}[Achievability]\label{achievability_thm}
	$\pi^{\textsf{MMW}}$ is a $2$-approximation scheduling policy for the following two scenarios:
	\begin{enumerate}
	\item static users located in a common cell 
	\item statistically identical users with i.i.d. uniform mobility. 
	\end{enumerate}
\end{theorem}
\end{framed}
Note that scenarios (1) and (2) represent two extreme ends of the user mobility landscape. Scenario (1) corresponds to zero mobility, whereas scenario (2) corresponds to an infinite mobility model in the limit. In our notations, the second scenario is characterized by $p_i=p, \forall 1\leq i \leq N$, $\psi_{ij}=\frac{1}{M}, \forall 1\leq i\leq N, 1\leq j \leq M,$ such that the cell occupancy random variables $\bm{C}_i(t)$'s are independent across time. The i.i.d. mobility model has been considered earlier in \cite{neely2005capacity} in connection with characterizing the capacity and delay tradeoffs in wireless networks.  
\begin{IEEEproof}
 Let the scheduling decisions at slot $t$ be denoted by the binary control vector $\bm{\mu}(t) \in \{0,1\}^N$, where $\mu_i(t)=1$ if and only if the following two conditions hold simultaneously: (1) $C_i(t)=j$, \emph{i.e.,} the $i$\textsuperscript{th} user is within the coverage area of the $j$\textsuperscript{th} BS at slot $t$, for some $1 \leq j \leq M$, and (2) $\textrm{BS}_j$ schedules a packet transmission  to the $i$\textsuperscript{th} user at time $t$. Since a BS can schedule only one transmission per slot to a user in its coverage area, the control vector must satisfy the following constraint: 
\begin{eqnarray*}
\sum_{i: C_i(t)=j}\mu_i(t) \leq 1, ~~ \forall j, t. 	
\end{eqnarray*}
 
%
For performance analysis, we consider the following linear Lyapunov function: 
\begin{eqnarray} \label{lyap_linear}
	L(\bm{h}(t))= \sum_{i=1}^N \frac{h_i(t)}{\sqrt{p_i}}.
\end{eqnarray}
The above linear Lyapunov function should be contrasted with the quadratic Lyapunov function used in \cite{kadota2018scheduling}. The conditional transition probabilities for the age of the $i$\textsuperscript{th} user may be expressed as follows (see Figure \ref{event_sequence}):
\begin{eqnarray*}
&&\mathbb{P}\big(h_i(t+1)=1|\bm{h}(t), \bm{\mu}(t+1), \bm{C}(t+1)\big) = \\
&&\mu_i(t+1)p_i,\\
&&\mathbb{P}\big(h_i(t+1)=h_i(t)+1|\bm{h}(t), \bm{\mu}(t+1), \bm{C}(t+1)\big) = \\
&&1 - \mu_i(t+1)p_i,
\end{eqnarray*}
where the first conditional probability corresponds to the event $\mathcal{E}_i(t)$ where the $i$\textsuperscript{th} user was scheduled at slot $t$ and the packet transmission was successful, and the second equation corresponds to the complement of the previous event. 
Hence, for each user $i \in [N]$, we have:
\begin{eqnarray} \label{one_step_expectation}
	\mathbb{E}\big(h_i(t+1)|\bm{h}(t), \bm{\mu}(t+1), \bm{C}(t+1)\big) \nonumber \\
	= h_i(t) -\mu_i(t+1)p_i h_i(t)+1.
\end{eqnarray}
 From the above equation, we can evaluate the one-step conditional drift of the Lyapunov function $L(\cdot)$ as follows:
 \begin{eqnarray} \label{drift_ineq_1}
  \mathbb{E}\big(L(\bm{h}(t+1))-L(\bm{h}(t)) | \bm{h}(t), \bm{\mu}(t+1), \bm{C}(t+1)\big) \nonumber\\
 = -\sum_{i=1}^N \mu_i(t+1) \sqrt{p_i}h_i(t) + \sum_{i=1}^N \frac{1}{\sqrt{p_i}}. 	
 \end{eqnarray}

Now consider the policy $\pi^{\textsf{MMW}}$, under which, each Base Station schedules the user $i$ having the highest weight $\sqrt{p_i}h_i(t)$ among all users in its cell. To analyze the performance of the $\pi^{\textsf{MMW}}$ policy, we define an auxiliary stationary randomized scheduling policy $\textsf{RAND}$, under which every BS randomly schedules a user in its cell with probability $ \mu^{\textrm{RAND}}_i(t+1) \propto 1/\sqrt{p_i}$ \footnote{We use the usual convention that summation over an empty set is zero.}. Using the basic fact that maximum of a set of real numbers is at least as large as any convex combination of the numbers, we conclude that:

\begin{eqnarray*}
	\mathbb{E}\bigg(\sum_{i=1}^N \mu^{\textsf{MMW}}_i(t+1) \sqrt{p_i}h_i(t)| \bm{h}(t), \bm{\mu}(t+1), \bm{C}(t+1)\bigg)\\
	\geq \sum_{j=1}^M \frac{\sum_{i: C_i(t+1)=j} h_i(t)}{\sum_{i: C_i(t+1)=j} \frac{1}{\sqrt{p_i}}}. 
\end{eqnarray*}
Hence, combining the above inequality with Eqn.\ \eqref{drift_ineq_1}, we have the following upper bound on the one-step expected drift of the Lyapunov function under the action of the policy $\pi^{\textsf{MMW}}:$
\begin{eqnarray} \label{drift_ub}
	\mathbb{E}^{\textsf{MMW}}\big(L(\bm{h}(t+1))-L(\bm{h}(t)) | \bm{h}(t), \bm{C}(t+1)\big) \nonumber \\
	\leq -\sum_{j=1}^M \frac{\sum_{i: C_i(t+1)=j} h_i(t)}{\sum_{i: C_i(t+1)=j} \frac{1}{\sqrt{p_i}}}+ \sum_{i=1}^N \frac{1}{\sqrt{p_i}}. 
\end{eqnarray}
Taking expectation of the above inequality w.r.t. the random cell-occupancy vector $\bm{C}(t+1)$, we have: 
\begin{eqnarray} \label{drift_mob}
	\mathbb{E}^{\textsf{MMW}}\big(L(\bm{h}(t+1))-L(\bm{h}(t))|\bm{h}(t)\big) \nonumber \\
	\leq -\sum_{j=1}^M \mathbb{E}(Z_j(t)|\bm{h}(t)) + \sum_{i=1}^N \frac{1}{\sqrt{p_i}},
\end{eqnarray}
where we define the random variable $Z_j(t) \equiv \frac{\sum_{i: C_i(t+1)=j} h_i(t)}{\sum_{i: C_i(t+1)=j} \frac{1}{\sqrt{p_i}}},\forall j,$ and let $0/0 \equiv 0.$ This convention is consistent, because if a cell is empty, the corresponding term is not present in the summation \eqref{drift_mob}.

 The inequality \eqref{drift_mob} holds for any arbitrary user mobility model. To make progress, we now analyze two special cases as stated in the statement of the theorem.


%
%
%
%
%

\textsc{Case I: Static users located in a common cell}

In this case, we have $M=1$, and hence, 
\begin{eqnarray*}
\mathbb{E}^{\textsf{MMW}}\big(L(\bm{h}(t+1))-L(\bm{h}(t))|\bm{h}(t)\big) 
	\leq - \frac{\sum_{i=1}^Nh_i(t)}{\sum_{i=1}^N\frac{1}{\sqrt{p}_i}} + \sum_{i=1}^N \frac{1}{\sqrt{p_i}}.
\end{eqnarray*}
Taking expectations of both sides w.r.t. $\bm{h}(t)$, we obtain:
\begin{eqnarray*}
	\mathbb{E}\big(L(\bm{h}(t+1))-L(\bm{h}(t))\big) \leq  - \frac{\sum_{i=1}^N\mathbb{E}(h_i(t))}{\sum_{i=1}^N\frac{1}{\sqrt{p}_i}} + \sum_{i=1}^N \frac{1}{\sqrt{p_i}}.
\end{eqnarray*}

Summing up the above inequalities for $t=1,2,\ldots, T$, dividing both sides by $T$ and then taking the limit as $T \to \infty$, we obtain:
\begin{eqnarray*}
\textsf{AoI}^{\textsf{MMW}}_{\textrm{avg}} = \limsup_{T \to \infty} \frac{1}{NT}\sum_{t=1}^T\sum_{i=1}^N \mathbb{E}(h_i(t))	
\leq  \frac{1}{N}\big(\sum_{i=1}^N \frac{1}{\sqrt{p}_i} \big)^2.
\end{eqnarray*}
Comparing the above with the lower bound in Eqn.\ \eqref{lb_expr} and realizing that $g(\psi)=1$ in this case, we have 
\begin{eqnarray*}
\textsf{AoI}^{\textsf{MMW}}_{\textrm{avg}} \leq 2 \textsf{AoI}_{\textrm{avg}}^*.	
\end{eqnarray*}

\textsc{Case II: Statistically identical users with i.i.d. uniform mobility}

In the case of i.i.d. uniform mobility, the users move to any one of the $M$ Base Stations chosen uniformly at random independent of everything else. We begin our analysis from the drift upper bound given in equation \eqref{drift_mob}. Note that, given the age vector $\bm{h}(t)$, under this mobility model, the r.v.s $Z_j(t)$'s are identically distributed. Hence, 
\[ \sum_{j=1}^M \mathbb{E}(Z_j(t)|\bm{h}(t))= M \mathbb{E}(Z_1(t)|\bm{h}(t)).\]
The RHS of the above can be expressed as the expectation of the ratio of two correlated random variables: 
\begin{eqnarray}\label{ratio_bd2}
	\mathbb{E}(Z_1(t)|\bm{h}(t)) = \mathbb{E}_{\bm{h}(t)}\bigg[\frac{\sum_{i=1}^Nh_i(t)W_i}{\sum_{i=1}^N\frac{1}{\sqrt{p_i}}W_i} \bigg], 
\end{eqnarray}
where the indicator r.v.s $\{W_i\}_{i=1}^N$ are i.i.d. such that 
\begin{eqnarray*}
\mathbb{P}(W_i=1)=\frac{1}{M}=1-\mathbb{P}(W_i=0).	
\end{eqnarray*}
In general, it is non-trivial to obtain a tight lower bound to the expression \eqref{ratio_bd2}, which is defined as the expectation of the ratio of two correlated random variables. In the following, we evaluate the expectation exactly in the special case of statistically identical users with $p_i=p,\forall i \in [N].$ 

Note that, we can express the random variable $\sum_{j=1}^M Z_j(t)$ as 
\begin{eqnarray*}
\sum_{j=1}^M Z_j(t) = \sum_{i=1}^N h_i(t) Y_i(t),	
\end{eqnarray*}
where $Y_i(t) = \big(\frac{1}{\sqrt{p_i}}+\sum_{k\neq i} \frac{1}{\sqrt{p_k}}\mathds{1}(C_i(t)=C_k(t))\big)^{-1}. $ 
%
%
%
%
%
%
%
Given our i.i.d. mobility assumption, the r.v. $\bm{C}(t+1)$ is independent of $\bm{h}(t)$. Hence,
\begin{eqnarray} \label{yi}
\mathbb{E}(Y_i(t)|\bm{h}(t)) = \sum_{n=0}^{N-1}\sum_{S: i\notin S, |S|=n} \bigg(\frac{1}{\sqrt{p_i}}+\sum_{k\in S} \frac{1}{\sqrt{p_k}}\bigg)^{-1} \times \nonumber \\
\frac{1}{M^n}\bigg(1-\frac{1}{M}\bigg)^{N-n-1}.
\end{eqnarray}
Since the users are statistically identical, \emph{i.e.,} $p_i=p, \forall i$, the summation \eqref{yi} has a closed-form expression. Clearly, for all $ 0 \leq n \leq N-1$, we have:
\begin{eqnarray*}
Y_i(t)= \frac{\sqrt{p}}{n+1},~~~ \textrm{w.p.}~ \binom{N-1}{n} \frac{1}{M^n}\bigg(1-\frac{1}{M}\bigg)^{N-n-1}.
\end{eqnarray*}
To evaluate the expectation of $Y_i(t)$, we integrate the binomial expansion of $(1+x)^{N-1}$ in the range $[0,\beta]$ to obtain the identity:
\begin{eqnarray*}
\frac{1}{N}\bigg( (1+\beta)^N-1 \bigg) = \beta \sum_{n=0}^{N-1} \frac{1}{n+1}\binom{N-1}{n} \beta^n. 	
\end{eqnarray*}
Substituting $\beta = \frac{1}{M-1}$ in the above, we obtain 
\begin{eqnarray}\label{iid_yi}
\mathbb{E}(Y_i(t)|\bm{h}(t))= \sqrt{p}\frac{M}{N}\bigg(1-\big(1-\frac{1}{M}\big)^{N} \bigg) \equiv Y^*(\textrm{say}). 	
\end{eqnarray}
From Eqn. \eqref{drift_mob} and \eqref{iid_yi}, we have 
\begin{eqnarray*}
\mathbb{E}^{\textsf{MMW}}\big(L(\bm{h}(t+1))-L(\bm{h}(t))|\bm{h}(t)\big) \leq -Y^*\sum_i h_i(t) + \frac{N}{\sqrt{p}}.	
\end{eqnarray*}
Taking expectation of both sides, we have 
\begin{eqnarray*}
	\mathbb{E}^{\textsf{MMW}}\big(L(\bm{h}(t+1))-L(\bm{h}(t))\big) \leq -Y^*\sum_i \mathbb{E}h_i(t) + \frac{N}{\sqrt{p}}.
\end{eqnarray*}
Summing up the above inequalities for $t=1,2,\ldots, T$, dividing both sides by $T$ and then taking limit as $T \to \infty$, we obtain
\begin{eqnarray}\label{ub_pf_1}
	\textsf{AoI}^{\textsf{MMW}}_{\textrm{avg}}=\limsup_{T \to \infty} \frac{1}{NT}\sum_{t=1}^{T}\sum_i \mathbb{E}h_i(t) \nonumber \\ 
	\leq \frac{N}{Y^*\sqrt{p}}= \frac{N}{Mp\bigg( 1- (1-\frac{1}{M})^N\bigg)}.
\end{eqnarray}
On the other hand, the lower bound from Theorem \ref{lb}, specialized to the case of identical users, yields:
\begin{eqnarray} \label{lb_pf_1}
\textsf{AoI}^*_{\textrm{avg}} \geq \frac{N}{2Mp\bigg( 1- (1-\frac{1}{M})^N\bigg)}.	
\end{eqnarray}
Eqns.\ \eqref{ub_pf_1} and \eqref{lb_pf_1}, we have 
\[\textsf{AoI}^{\textsf{MMW}}_{\textrm{avg}} \leq 2\textsf{AoI}_{\textrm{avg}}^*. \]
The above inequality shows that the $\textsf{MMW}$ scheduling policy is $2-$optimal for statistically identical users with uniform i.i.d. mobility.
\end{IEEEproof}
\paragraph*{Discussion} \editr{In this Section, we have presented a universal lower bound (Theorem \ref{lb}) and near-optimal achievability results for two important special cases (Theorem \ref{achievability_thm}) for the average AoI metric in the stochastic setting. It will be interesting to establish similar achievability results for more general user mobility (\emph{e.g.,} random walk) and channel models (\emph{e.g.,} Gilbert-Elliot model \cite{gilbert-AoI})}.

\subsection{Minimizing the Peak AoI in the Stochastic Framework}\label{max_stochastic}
In this section, we consider the problem of minimizing the long-term \emph{peak}-AoI metric \eqref{max_objective}, for static users in a single cell. By directly solving the associated countable state Bellman equation, we prove that the greedy \textsf{CMA} scheduling policy is optimal for minimizing the Peak AoI. Furthermore, we also establish the large-deviation optimality of the \textsf{CMA} policy. 
\begin{framed}
\begin{theorem}[Optimal Policy for minimizing the Peak AoI] \label{opt}
	The greedy \textsf{CMA} policy is optimal  for minimizing the peak AoI objective given in Eqn.\ \eqref{max_objective}. Moreover, the optimal value of the peak AoI is given by $\textrm{AoI}_{\textrm{peak}}^*=\sum_{i=1}^{N} \frac{1}{p_i}$.
\end{theorem}
\end{framed}
We prove Theorem \ref{opt} by proposing a closed-form solution to the Bellman's equation corresponding to the associated average-cost MDP and then verifying that the candidate solution indeed satisfies the Bellman's equation.
 \begin{IEEEproof}
 	The stochastic control problem under investigation is an instance of a countable-state average-cost MDP with a finite action space with the following components. The state of the system at a slot $t$ given by the instantaneous AoI of all users: $\bm{h}(t)\equiv (h_1(t),h_2(t), \ldots, h_N(t))$. The per-stage cost at time $t$ is $\max_{i=1}^{N}h_i(t)$, which is unbounded, in general. Finally, the finite action space $\mathcal{A}=\{1,2,\ldots, N\}$ corresponds to the user scheduled at a given slot. \\
 	Let the optimal cost for the problem  be denoted by $\lambda^*$ and the differential cost-to-go from the state $\bm{h}$ be denoted by $V(\bm{h})$ \cite{bertsekas2005dynamic}. Then, following the standard theory of average cost countable state MDP (Proposition 4.6.1 of \cite{bertsekas2005dynamic}), we set up the following Bellman Eqn.  
 	\begin{eqnarray} \label{bellman}
 		&&\lambda^*+ V(\bm{h})\\
 		&=& \min_i\{ p_i V(\underbrace{1}_{i\textsuperscript{th} \textrm{ coordinate}}, ~\bm{h}_{-i}\bm{+1})+ (1-p_i) V(\bm{h+1})\} \nonumber \\
 		&+& \max_i h_i \nonumber, 
 	\end{eqnarray}
 	where the vector $\bm{h}_{-i}$ denotes the $N-1$ dimensional vector of all coordinates excepting the $i$\textsuperscript{th} coordinate and $\bm{1}$ is an all-one vector of an appropriate dimension.
 	\paragraph*{Discussion} The Bellman Equation \eqref{bellman} may be derived as follows. Suppose that the current age of the users is given by the vector $\bm{h}$. If the policy schedules a packet transmission to the $i$\textsuperscript{th} user, the transmission is successful with probability $p_i$ and is unsuccessful with probability $1-p_i$. If the transmission is successful, the AoI of all users, excepting the $i$\textsuperscript{th} user, is incremented by $1$, and the AoI of the $i$\textsuperscript{th} user is reduced to $1$. This explains the first term. On the other hand, if the transmission to the $i$\textsuperscript{th} user is unsuccessful, the AoI of all users are incremented by $1$. This explains the second term within the bracket. Finally, the term $\max_i h_i$ denotes the stage cost.
 	   
\paragraph*{Solution to the Bellman Equation \eqref{bellman}} We now verify that the following constitutes a solution to the Bellman Equation \eqref{bellman}:
\begin{eqnarray} \label{bellman_soln}
	V(\bm{h})= \sum_j \frac{h_j}{p_j}, \hspace{10pt}
	\lambda^*=\sum_j \frac{1}{p_j}.
\end{eqnarray} 	
 To verify the above solution, we start with the RHS of Eqn.\ \eqref{bellman}. Upon substitution from Eqn. \eqref{bellman_soln}, the expression corresponding to the $i$\textsuperscript{th} user inside the $\min$ operator of Eqn. \eqref{bellman} simplifies to: 
 \begin{eqnarray}
 	&&p_i V(1, \bm{h}_{-i}\bm{+1})+ (1-p_i) V(\bm{h+1}) \nonumber \\
 	&=& p_i \sum_{j\neq i} \frac{h_j+1}{p_j}+ 1 + (1-p_i) \sum_{j=1}^N \frac{h_j+1}{p_j}\nonumber \\
 	&=& p_i \sum_{j=1}^N \frac{h_j+1}{p_j} - p_i\frac{h_i+1}{p_i}+1 + (1-p_i) \sum_{j=1}^N \frac{h_j+1}{p_j}\nonumber \\
 	&=& \sum_{j=1}^N \frac{h_j}{p_j} - h_i + \sum_{j=1}^N \frac{1}{p_j}. \label{calc}
  \end{eqnarray}	
  Hence,
  \begin{eqnarray*}
  && \textsf{RHS}\\
  	&=&\min_i\{ p_i V(1, \bm{h}_{-i}\bm{+1})+ (1-p_i) V(\bm{h+1})\} + \max_i h_i \\
  	&=& \sum_{j=1}^N \frac{1}{p_j} + \sum_{j=1}^N\frac{h_j}{p_j} - \max_i h_i + \max_i h_i\\
  	&=& \lambda^* + V(\bm{h})\\
  	&=& \textsf{LHS}.
  \end{eqnarray*}
 Finally, to verify the regularity condition (Eqn.\ $4.122$ of \cite{bertsekas2005dynamic}), note that
 \begin{eqnarray*}
 \frac{\mathbb{E}(V(\bm{h}(t))| \bm{h}(1))}{t}&=& \frac{1}{t}\mathbb{E}\bigg[\sum_{j=1}^N \frac{h_j(t)}{p_j}|\bm{h}(1)\bigg]	\\
 &\leq & \mathbb{E}\bigg[\frac{\max_{i=1}^N h_i(t)}{t}| \bm{h}(1)\bigg](\sum_i \frac{1}{p_i}).
 \end{eqnarray*}
 Hence, for any scheduling policy $\pi$ for which the regularity condition is violated, \emph{i.e.,} $\limsup_{t \to \infty}  \frac{\mathbb{E}^\pi(V(\bm{h}(t))| \bm{h}(1))}{t} >0,$ for some $\bm{h}(1)$ with positive probability, the above bound implies that
 \begin{eqnarray*}
 \limsup_{T \to \infty} \frac{1}{T}\sum_{t=1}^{T}\mathbb{E}^\pi \big[\max_{i=1}^N h_i(t)\big]	= \infty.
 \end{eqnarray*}
 Hence, without any loss of optimality, we may confine our attention to those policies for which the regularity condition holds. The proof now follows from Proposition $4.6.1$ of \cite{bertsekas2005dynamic}.
\end{IEEEproof} 
\subsubsection*{Large-Deviation Optimality} 
Theorem \ref{opt} establishes that the greedy \textsf{CMA} scheduling policy is optimal for minimizing the long-term \emph{expected} peak AoI. However, for some applications, the scheduling policy is additionally required to ensure that the peak AoI metric stays within limit with a high probability after a sufficiently long time. An analogous problem was considered in connection with the queue stability in a similar multi-user single cell setting in \cite{ying2006large}.
The previous requirement leads to the following problem statement: Design a scheduling policy $\pi \in \Pi$ that maximizes the large deviation exponent, \emph{i.e.,}
\begin{eqnarray} \label{ld_problem} 
 \max_{\pi \in \Pi }\bigg[-\lim_{k\to \infty}\lim_{t\to \infty}\frac{1}{k}\log \mathbb{P}^{\pi}(\max_{i=1}^N h_i(t) \geq k)\bigg].
 \end{eqnarray} 
The following theorem shows that the \textsf{CMA} policy is also optimal in the above sense.
\begin{framed}
\begin{theorem}\label{ld_opt}
The \textsf{CMA} policy achieves the optimal large-deviation exponent whose value is given by 
	\begin{equation*}
		-\lim_{k\to \infty}\lim_{t\to \infty}\frac{1}{k}\log \mathbb{P}^{\textsf{MA}}(\max_i h_i(t) \geq k) = - \log(1-\min_{i=1}^Np_{i}).
	\end{equation*}	
\end{theorem}
\end{framed}
See Appendix \ref{ld_opt_proof} for the proof.

\section{Numerical Experiments}\label{simulation}
\begin{figure}
\centering	
\begin{overpic}[width=0.3\textwidth]{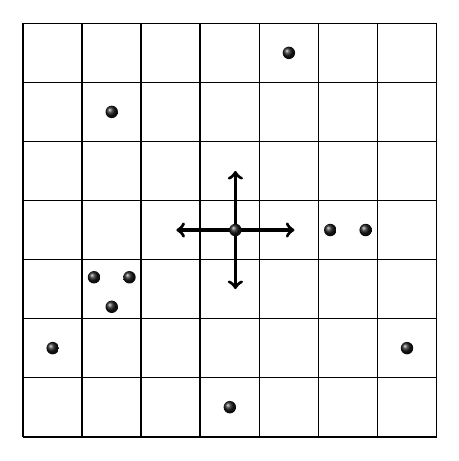}
\end{overpic}
\caption{Schematic of a grid network used in the simulations \edit{in the stationary regime}. The black balls represent the users and the small squares denote the cells. There is one Base Station at every cell (not shown in this schematic). At every step, each user randomly moves to any one of the adjacent cells with equal probability.}
\label{sim_setup}
\end{figure}

\subsection{Stationary regime} In our numerical experiments, we first simulate the $\textsf{CMA}$ and the \textsf{MMW} policies  \edit{in the stationary regime} and compare their performances for minimizing the average AoI and peak AoI.  We consider a grid network with $M=100$ base stations arranged in the form of a $10 \times 10$ square grid. $N$ users are initially placed uniformly at random on any of the $M$ Base Stations.  Each user executes an independent 2-dimensional random walk on the square grid at every time slot and moves to any one of the adjacent cells with equal probability. See Figure \ref{sim_setup} for a schematic. As in our system model, we assume that each Base Station can schedule a new packet transmission to only one user located in its cell per slot. \edit{In the stationary regime}, the transmission success probabilities $\bm{p}$ of each of the users are sampled independently and uniformly at random from the interval $I  \subseteq [0,1]$ and remains fixed throughout the simulation duration.
 
\begin{figure}
\centering
\hspace{-10pt}
\begin{overpic}[width=0.45\textwidth]{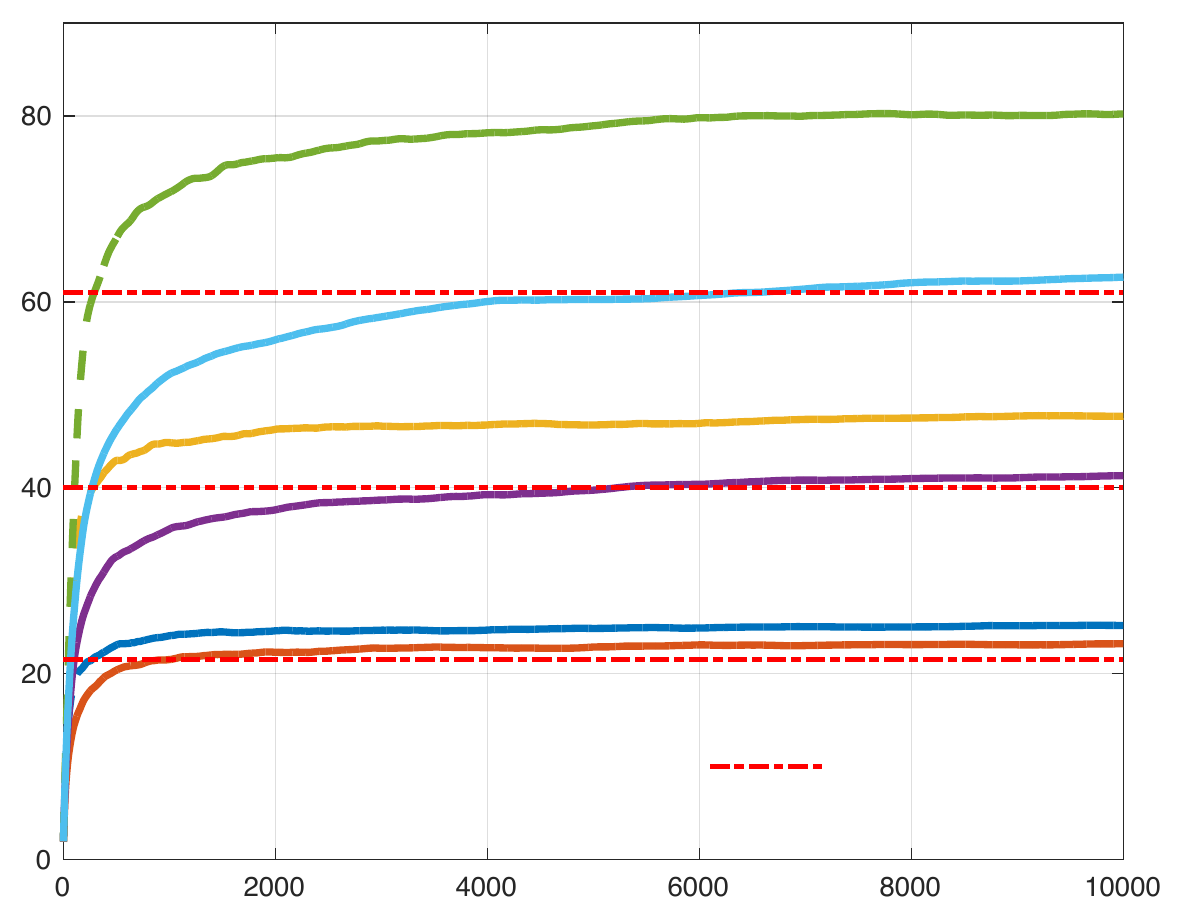}
\put(72,18){\footnotesize{$N=10^3$}}
\put(70,39){\footnotesize{$N=2\times 10^3$}}
\put(70,60){\footnotesize{$N=3\times 10^3$}}
\put(67, 63){\footnotesize{$\nwarrow$}}
\put(67, 56){\footnotesize{$\swarrow$}}
\put(30,-2){\footnotesize{Simulation duration (in slots)}}
\put(-4,15){\rotatebox{90}{\footnotesize{Time-averaged Average AoI (in slots)}}}
\put(12,15){$\boldsymbol{-}\boldsymbol{-}\boldsymbol{-}$}
\put(23,15){\footnotesize{\textsf{CMA}}}
\put(55,9){\footnotesize{\textsf{Lower Bound}}}
\put(23,10){\footnotesize{\textsf{MMW}}}
\put(13,10){$\textbf{------}$}
\end{overpic}
\caption{Comparison between the \textsf{CMA} and \textsf{MMW} policy for statistically non-identical users for the Average-age metric \edit{in the stationary regime}. The users are assumed to execute independent 2-D random walks on a $10 \times 10$ grid. The transmission success probability of each user is sampled uniformly at random from the interval $[0,1]$. From the plots, it follows that the \textsf{MMW} policy outperforms the \textsf{CMA} policy for minimizing the average AoI. }
\label{comp_fig_Avg-age}
\end{figure}

\begin{figure}
\centering
\hspace{-10pt}
\begin{overpic}[width=0.45\textwidth]{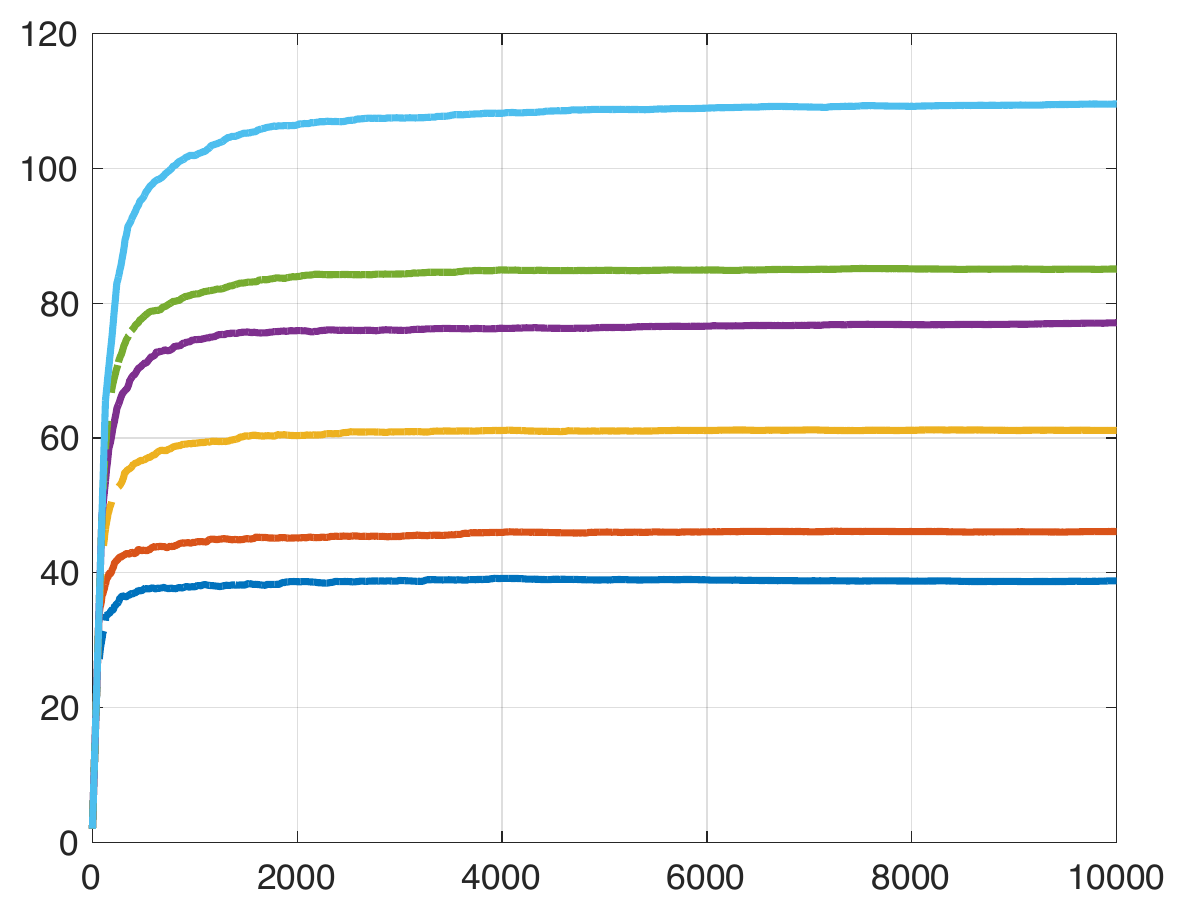}
\put(72,23){\footnotesize{$N=10^3$}}
\put(70,44){\footnotesize{$N=2\times 10^3$}}
\put(70,60){\footnotesize{$N=3\times 10^3$}}
\put(35,-2){\footnotesize{Simulation duration (in slots)}}
\put(-2,15){\rotatebox{90}{\footnotesize{Time-averaged Peak AoI (in slots)}}}
\put(12,17){$\boldsymbol{-}\boldsymbol{-}\boldsymbol{-}$}
\put(67, 65){\footnotesize{$\nwarrow$}}
\put(67, 56){\footnotesize{$\swarrow$}}
\put(67, 47){\footnotesize{$\nwarrow$}}
\put(67, 42){\footnotesize{$\swarrow$}}
\put(23,17){\footnotesize{\textsf{CMA}}}
\put(23,12){\footnotesize{\textsf{MMW}}}
\put(13,12){$\textbf{------}$}
\end{overpic}
\caption{Comparison between the \textsf{CMA} and \textsf{MMW} policy for statistically non-identical users for the Peak-Age metric \edit{in the stationary regime}. The users are assumed to execute independent 2-D random walks on a $10 \times 10$ grid. The transmission success probability of each user is sampled uniformly at random from the interval $[0.2,0.8]$. From the plots, it follows that the \textsf{CMA} policy outperforms the \textsf{MMW} policy for minimizing the peak AoI. }
\label{comp_fig_Max-age}
\end{figure}
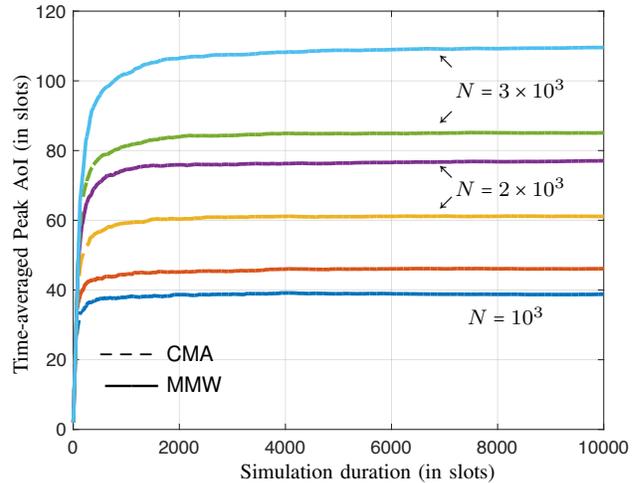

\paragraph*{Results} Figure \ref{comp_fig_Avg-age} shows the time variation of the average AoI for three different scenarios $N=1000, 2000, \textrm{and } 3000.$ Not surprisingly, the average AoI increases as the number of users $N$ in the network is increased. More interestingly, by numerically evaluating the lower bound \eqref{lb_expr} using the values of the simulation parameters, we see that the simulated long-term AoI under the \textsf{MMW} policy is very close (within $\sim 1\%$) to the theoretical lower bound.  This result indicates that (1) the lower bound is tight, and (2) the performance of the \textsf{MMW} policy is near-optimal. We also see that the greedy \textsf{CMA} policy is outperformed by the \textsf{MMW} policy, which takes into account the transmission success probabilities of different users. However, as the plot \ref{comp_fig_Max-age} shows, when it comes to maximizing the peak AoI,  the greedy \textsf{CMA} policy consistently outperforms the \textsf{MMW} policy. Hence, the choice of the particular AoI metric plays an important role in determining the performance of different scheduling policies.

\begin{figure}
\centering
\hspace{-10pt}
\begin{overpic}[width=0.45\textwidth]{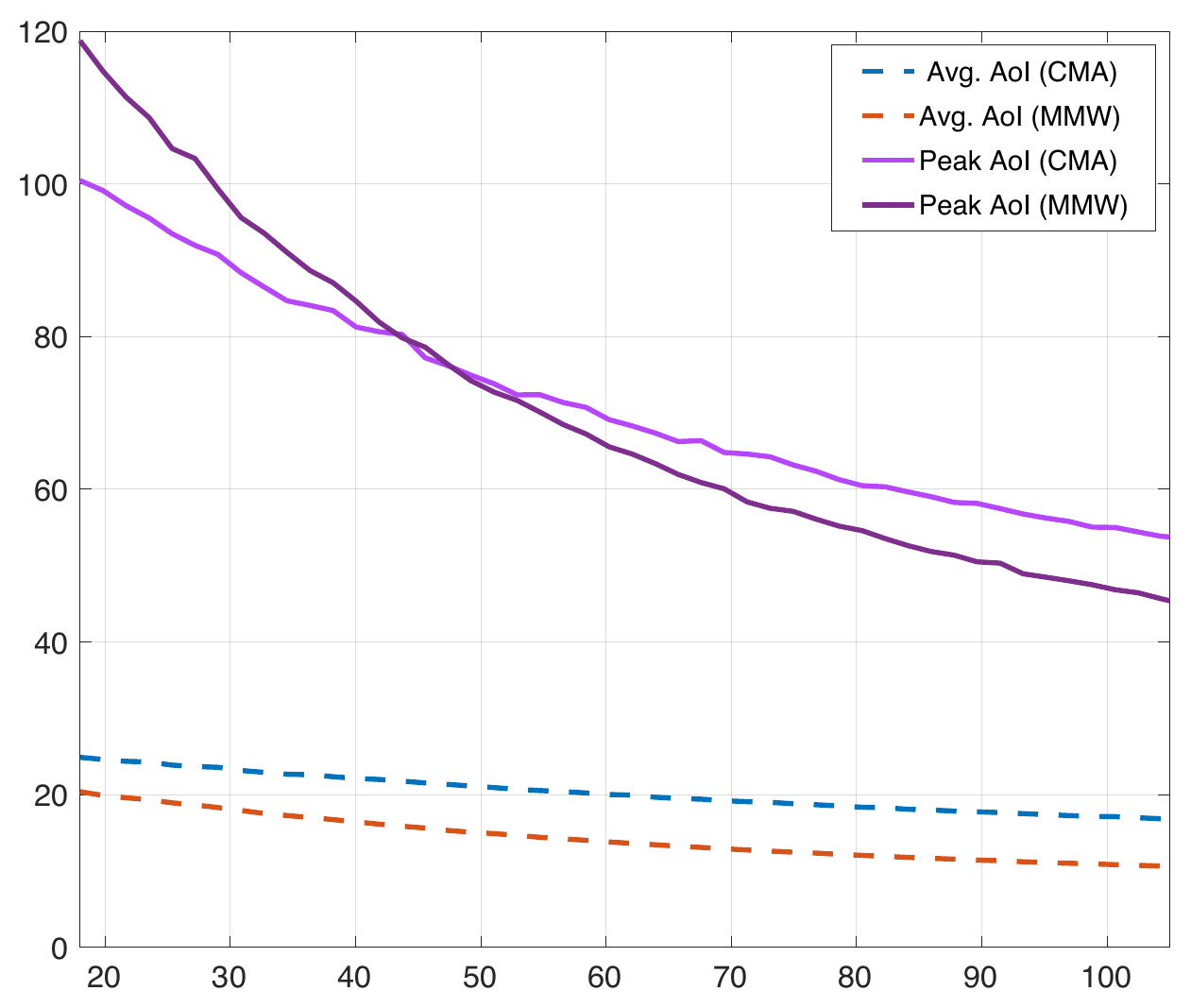}	
\put(45,-2){\footnotesize{$v_{\max}$ (in $\textrm{km/hr}$) }}
\put(-2,30){\rotatebox{90}{\footnotesize{AoI (in seconds)}}}
\end{overpic}
\caption{\edit{Performance comparison between the \textsf{CMA} and \textsf{MMW} policies in the non-stationary regime. The propagation environment corresponds to an NLOS Urban Microcell environment consisting of $100$ Base Stations and $N=1000$ users. From the plots, we see that AoI decreases as the mobility of the users increases. Furthermore, the \textsf{MMW} policy out-performs the \textsf{CMA} policy in the high-mobility regime.}}
\label{AoI-comp}
\end{figure}

\edit{
\subsection{Non-stationary regime} Next, we numerically evaluate the performance of the proposed scheduling policies in a non-stationary setup where the probability of successful transmission to each user is allowed to vary with time. Specifically, in this set of simulations, the users' movement is modeled using the \emph{L\'evy Mobility model}, which is known to capture human mobility accurately \cite{rhee2011levy}. We simulate a discrete time version of this model, where at each slot, each user moves independently at random by a distance $\rho$ at an angle $\theta$ from its current location. The distance $\rho$ and the angle $\theta$ are sampled uniformly at random from the intervals $[0, v_{\max}]$ and $[0, 2\pi]$ respectively, where $v_{\max}$ is a constant determined by the maximum velocity of the users. We assume that a packet loss occurs when the received SNR falls below a certain threshold (\emph{i.e.,} when a channel \emph{outage} occurs). Under this assumption, we use the following approximate expression for the probability of a successful packet reception over a slow-fading wireless channel for a user $i$ located at a distance $d_i(t)$ from its associated BS \cite{tse2005fundamentals}: 
\begin{align} \label{p-i-eqn}
p_i(t) &= 1-\bigg(\frac{1}{1+\underbrace{G\cdot L(d_i(t))^{-1} \cdot \textsf{SNR}_i}_{\text{received SNR}}} \bigg)^\eta
\end{align}
 In the above expressions, $G$ is the receive antenna gain, $L(\cdot)$ denotes the path loss function, $\eta$ is the diversity order of the system, and $\textsf{SNR}_i$ represents the transmit Signal-to-Noise ratio for the $i$\textsuperscript{th} user. Note that the received SNR for a user is obtained by multiplying the transmit SNR, receive-antenna gain, and the reciprocal of the path loss function. In our simulations, we use the following mean alpha-beta-gamma (ABG) 5G path loss model as proposed in \cite{sun2016propagation}:
 \begin{eqnarray} \label{loss-eqn}
 	L^{\textrm{ABG}}(d) [\textrm{dB}] = 10 \alpha \log_{10}\bigg(\frac{d}{1~\textrm{m} } \bigg) + \beta + 10\gamma \log_{10} \bigg(\frac{f_c}{1~ \textrm{GHz}} \bigg).
 \end{eqnarray}
 In the path loss formula above, $\alpha$ corresponds to the path-loss exponent, $\beta$ is an optimized offset value for the path loss in dB and the parameter $\gamma$ gives the dependence of the path loss on the carrier frequency $f_c$.
\paragraph*{Simulation setup} We consider a 5G mmWave cellular wireless system operating at a carrier frequency of $f_c=28~\textrm{GHz}.$ Assume that $M=100$ base stations are arranged in the form of a $10 \times 10$ uniform square grid such that the Inter-Site Distance (ISD) between any two adjacent base stations is $100 ~\textrm{m}.$ Base stations periodically receive fresh update packets every second from external sources, and each BS schedules a downlink transmission  immediately upon the reception of a fresh packet. We assume that $N=1000$ users are roaming in the area at the maximum speed of $v_{\max}~ \textrm{km/hr}$ according to the L\'evy mobility model described above. At the beginning of the simulations, the users are placed uniformly at random over the entire region. The path-loss parameters are taken to be $\alpha = 3.5, \beta = 24.4 ~\text{dB}, \gamma = 1.9,$ which correspond to an NLOS Urban Microcell environment \cite{sun2016propagation}. The diversity order of the system is taken to be $\eta = 5$, the transmit Signal-to-Noise ratio (\textsf{SNR}) is taken to be $90 ~\textrm{dB},$ and the gain of the receive antenna is taken to be $G= 20~\textrm{dB}$. 	
\paragraph*{Results and Discussions} We numerically compare the performance of the following two policies in the non-stationary regime - (1) the \textsf{CMA} policy, and (2) the \textsf{MMW} policy. The \textsf{MMW} policy updates the probability of successful transmission to each user according to its distance from the associated BS (viz.\ Eqns.\ \eqref{p-i-eqn} and \eqref{loss-eqn}). The average and the peak AoI metrics attained by these two policies are plotted in Figure \ref{AoI-comp} as a function of the maximum speed of the users $v_{\max}$. From the plots, we see that, in general, the peak and average AoI decrease as the maximum speed of the users increases. Furthermore, the \textsf{MMW} policy performs better than the \textsf{CMA} policy in minimizing the average AoI metric. On the other hand, the \textsf{CMA} policy is more effective in reducing the peak AoI metric up to some critical speed, beyond which the \textsf{MMW} policy performs better.    
 
}

%

%
\section{Concluding Remarks} \label{conclusion}
In this paper, we investigated the fundamental limits of Age-of-Information for mobile users in adversarial and stochastic environments. We also proposed efficient scheduling policies that come close to achieving the limits. In particular, we showed that a greedy scheduling policy (\textsf{CMA}) is near-optimal for minimizing the peak AoI in the adversarial setting. The competitive ratio of the same policy is shown to be within a factor of $O(N)$ of the optimal value for minimizing the average AoI.   
In the stochastic setting, we proved that a Max-Weight-type scheduling policy (\textsf{MMW}) attains $2$-approximation for minimizing the average AoI in two extreme mobility scenarios. Furthermore, the \textsf{CMA} policy is shown to be optimal for minimizing the peak AoI for static users in a single cell. Although, in this paper, we made some progress in the AoI-optimal scheduling problem for cellular networks, many interesting questions are still open. \editr{In the adversarial setting, an interesting problem is to reduce the current $O(N)$ gap to optimality for minimizing the average AoI metric.} The problem of designing an optimal policy for minimizing the peak AoI for mobile users in the stochastic environment is open and can be investigated in the future. \editr{Extending the achievability result in the stochastic setting to more general mobility and channel state models (\emph{e.g.,} Gilbert-Elliot model \cite{gilbert-AoI}) is an important research direction.} Furthermore, designing optimal scheduling algorithms in the adversarial setting when some estimate of the channel state for the immediate future is known, constitutes an interesting problem from a practical point-of-view \cite{aoi-est}. 

\section*{Acknowledgements} \label{ack_section}
This work is partially supported by the grant IND-417880 from
Qualcomm, USA and a research grant from the Govt.\ of India under the Institutes of Eminence (IoE) initiative. 
The first author would like to thank his former students Subhankar Banerjee and Arunabh Srivastava for some useful discussions during the early stages of this work.  


%

\section{Appendix} \label{appendix}

\subsection{Proof of Theorem \ref{tightness_thm}}\label{tightness_thm_proof}

\begin{IEEEproof}
We prove the theorem by exhibiting a channel state sequence for which the \textsf{CMA} policy achieves a competitive ratio of at least $N^2$ for the average AoI objective and at least $2N-1$ for the peak AoI objective.\\
 Consider a single-cell scenario where a BS serves $N$ stationary users. In this case, it is easy to see that the \textsf{CMA} policy reduces to a persistent round-robin policy - the users are scheduled in a round-robin fashion such that each user is scheduled continuously until its transmission is successful. Consider the following channel-state sequence: all super-intervals are of constant length $\Delta \gg N,$ where we take $\Delta \equiv 1 \mod (N-1)$. During each super-interval, all users, apart from the \textsf{Max}-user, have \textsf{Good} channels at every slot.
  Hence, under the \textsf{CMA} policy, the time interval between two consecutive successful packet transmissions to any user is $N\Delta$ slots. Thus, at the beginning of any super-interval (apart from the first $N-1$ super-intervals) the ages of the users under the \textsf{CMA} policy in \emph{ascending} order is given by $1,\Delta+1, 2\Delta+1,....,(N-1)\Delta+1$. Next, we consider the following two objectives.
 
 \paragraph{Average AoI}  The total cost incurred by the \textsf{CMA} policy in any interval (apart from the first $N-1$ super-intervals) is given by:
\begin{eqnarray*}
	C_i^{\textsf{CMA}} = \sum_{j=1}^N \sum_{k=1}^\Delta \big( (j-1)\Delta + k\big)= \frac{1}{2}\big(N^2\Delta^2+ N\Delta).
\end{eqnarray*}

Next, we upper-bound the cost incurred by the offline optimal policy \textsf{OPT} by comparing it to another (potentially sub-optimal) offline policy $\mathcal{P}$. The policy $\mathcal{P}$ serves each of the $N-1$ users other than the \textsf{Max} user in a round-robin fashion in each super-interval and finally, it serves the \textsf{Max} user at the last time slot of each super-interval (See Figure \ref{intervals_fig}). Clearly, under the action of the policy $\mathcal{P},$ the set of ages of the users at the beginning of every super-interval is given by $\{1,2,\ldots, N\}.$
Thus the total cost incurred by the policy $\mathcal{P}$ during the $i$\textsuperscript{th} super-interval is bounded as:
\begin{eqnarray}\label{opt_seq}
 C_i^{\mathcal{P}} &\leq & \underbrace{\sum_{k=1}^{\Delta} k}_{\textrm{Cost due to the \textsf{Max} user}} \nonumber \\
 &&+ \underbrace{(N-1)\bigg(\big(\frac{\Delta-1}{N-1} -1 \big) \frac{N(N-1)}{2} +\frac{3N(N-1)}{2}\bigg)}_{\textrm{Cost due to the other }N-1 \textrm{ users}} \nonumber \\
&\leq & \frac{\Delta^2}{2}+\frac{\Delta N^2}{2} + N^3.
\end{eqnarray}
Let $K$ be the number of super-intervals in the time-horizon $T$. We have 
\begin{eqnarray*}
\textsf{Cost}^{\textsf{OPT}}(T)  \leq \sum_{i=1}^K 	C_i^{\mathcal{P}}.
\end{eqnarray*}
Hence, for large enough $K$, the competitive ratio of the \textsf{CMA} policy is lower bounded as: 
\begin{eqnarray*}
\eta^{\textsf{CMA}}_{\textrm{avg}} = \frac{\sum_{i=1}^KC_i^{\textsf{CMA}}}{\sum_{i=1}^K C_i^{\textsf{OPT}}} \geq \frac{\frac{1}{2}(N^2\Delta^2+N\Delta)}{\frac{\Delta^2}{2}+\frac{\Delta N^2}{2} + N^3}.	\end{eqnarray*}
By taking $\Delta$ to be arbitrarily large, it follows from the above expression that 
\begin{eqnarray*}
	\eta^{\textsf{CMA}}_{\textrm{avg}} \geq N^2.
\end{eqnarray*}

\paragraph{Peak AoI} We compute the competitive ratio for the same channel state sequence as before for the peak AoI objective. The cost incurred by the \textsf{CMA} policy during any super-interval is:
\begin{equation}
    C_i^{\textsf{CMA}} =\sum_{k=1}^\Delta \bigg( (N-1)\Delta + k \bigg) =  (N-1)\Delta^2 + \frac{\Delta(\Delta+1)}{2}.
\end{equation}
Similar to the Average AoI case, we use the policy $\mathcal{P}$ to upper-bound the cost incurred by the \textsf{OPT} policy for the given channel state sequence. As before, the total cost incurred by the policy $\mathcal{P}$ during the $i$\textsuperscript{th} super-interval is bounded as:
\begin{eqnarray*}
C_i^{\mathcal{P}} \leq 	\sum_{k=1}^{\Delta} (N+k) = N\Delta + \frac{\Delta(\Delta+1)}{2}.
\end{eqnarray*}

Thus, summing over all super-intervals, we obtain
\begin{eqnarray*}
\eta^{\textsf{CMA}}_{\textrm{peak}}	= \frac{\sum_{i=1}^KC_i^{\textsf{CMA}}}{\sum_{i=1}^K C_i^{\textsf{OPT}}} \geq \frac{(N-1)\Delta^2 + \frac{\Delta(\Delta+1)}{2}}{N\Delta + \frac{\Delta(\Delta+1)}{2}}.
\end{eqnarray*}
The result now follows by letting the length of the sub-intervals $\Delta$ to be arbitrarily large.

\end{IEEEproof}

\subsection{Proof of Proposition \ref{improved_LB}} \label{improved_LB_proof}
\begin{IEEEproof}
We use the same proof technique as in Theorem \ref{comp_ratio_lb} with the channel states for the two user being i.i.d. $\{\textsf{Good}, \textsf{Bad}\}$ or $\{\textsf{Bad}, \textsf{Good}\}$ w.p. $\nicefrac{1}{2}$ each. 
Let $\vec{\mu}(t) \in \{ \begin{pmatrix}1 & 0 \end{pmatrix}, \begin{pmatrix} 0 & 1 \end{pmatrix}\}$ denote the scheduling decision of the policy $\pi$ at slot $t$.  
Define $\mathcal{F}_{t-1}\equiv \sigma(\vec{h}(k), \vec{\mu}(k), 1\leq k \leq t-1)$ to be the sigma-algebra generated by the age and scheduling decision r.v.s up to time $t-1$. For any online policy, the scheduling decision $\vec{\mu}(t)$ at time $t$ must be measurable in $\mathcal{F}_{t-1}, \forall t\geq 1$.  Let $H_{\textrm{sum}}(t) \equiv \mathbb{E}^\pi(h_1(t))+ \mathbb{E}^\pi(h_2(t))$ be the expected sum of the ages of the users at time $t$. Let $B_t \in \mathcal{F}_t$ be the event for which the user $1$ is scheduled under the policy $\pi$. Then, we can write 
\begin{eqnarray} \label{cond_ex1}
&&\mathbb{E}^\pi\big(h_1(t+1)|\mathcal{F}_t) \\
&=& \big(1+\frac{1}{2}h_1(t)\big)\mathds{1}(B_t) + \big(1+h_1(t\big) \mathds{1}(B_t^c) \nonumber \\
 &=& 1+ \frac{1}{2}h_1(t) + \frac{1}{2}h_1(t)\mathds{1}(B_t^c)\nonumber \\
 &\stackrel{(a)}{\geq} & 1+ \frac{1}{2}h_1(t) + \frac{1}{2}\min \{h_1(t), h_2(t)\}\mathds{1}(B_t^c),
\end{eqnarray}
Similarly, we can also write
\begin{eqnarray}\label{cond_ex2}
	&&\mathbb{E}^\pi\big(h_2(t+1)|\mathcal{F}_t) \geq \nonumber \\
	&& 1+ \frac{1}{2}h_2(t) + \frac{1}{2}\min\{h_1(t), h_2(t) \}\mathds{1}(B_t).
\end{eqnarray}
Since $\mathds{1}(B_t)+\mathds{1}(B_t^c)=1$, from the equations \eqref{cond_ex1} and \eqref{cond_ex2}, we have 
\begin{eqnarray*}
	&&\mathbb{E}^\pi\big(h_1(t+1)+h_2(t+1)|\mathcal{F}_t) \geq \\
	 && 2+ \frac{1}{2}(h_1(t)+h_2(t))+ \frac{1}{2} \min\{h_1(t), h_2(t)\}.
\end{eqnarray*}
Taking expectations of both sides of the above equation, we get
\begin{eqnarray} \label{key_eqn1}
H_{\textrm{sum}}(t+1) \geq 2 + \frac{1}{2} H_{\textrm{sum}}(t) + \frac{1}{2} \mathbb{E}\bigg(\min\{h_1(t), h_2(t)\}\bigg).	
\end{eqnarray}
Let the random variable $S(t)$ denote the time elapsed since the last successful transmission (by any user) before time $t$. Clearly, \[ \min \{h_1(t), h_2(t)\} \geq S(t)\] (the above inequality holds with equality for the two user case).
Hence, the above inequality implies 
\[ H_{\textrm{sum}}(t+1) \geq 2 + \frac{1}{2} H_{\textrm{sum}}(t) + \frac{1}{2} \mathbb{E}\big(S(t)\big).\]
Summing up the above inequalities for $t=1,2, \ldots, T$, and dividing both sides by $T$, we obtain 
\begin{eqnarray} \label{cesaro_mean_lt}
	2\frac{H_{\textrm{sum}}(T+1)}{T}+\frac{1}{T}\sum_{t=1}^{T}H_{\textrm{sum}}(t) \geq 4 + \frac{1}{T}\sum_{t=1}^{T}\mathbb{E}(S(t)).
\end{eqnarray}
  It is to be noted that $\{S(t)\}_{t\geq 1}$ is a renewal process with the time-stamp of successful transmissions constituting the renewal instants. Let the random variable $\tau$ denote the length of any generic renewal cycle. Hence, using the renewal reward theorem \cite{gallager2012discrete} \cite{gallager2013stochastic}, it follows that 
\begin{eqnarray*}
	\lim_{T \to \infty} \frac{1}{T}\sum_{t=1}^{T}\mathbb{E}(S(t)).&=& \frac{\mathbb{E}\big(\int_{0}^{\tau}S(t)dt\big)}{\mathbb{E}(\tau)}\\
	&=& \frac{\mathbb{E}(1+2+\ldots+\tau)}{\mathbb{E}(\tau)}\\
	&=& \frac{\mathbb{E}(\tau^2)+\mathbb{E}(\tau)}{2\mathbb{E}(\tau)}\\
	&=& 2,
\end{eqnarray*}

where the last inequality follows from the fact that the renewal cycle lengths $T$ are distributed geometrically with the parameter $p=1/2$. Thus, the limit of the RHS of Eqn.\ \eqref{cesaro_mean_lt} exists and the limiting value is equal to $6$.
Next, we consider two possible cases. \\
\textbf{Case I: $\liminf_{T\to \infty} \frac{H_{\textrm{sum}}(T+1)}{T}=0$: } In this case, consider a subsequence $\{T_k\}_{k\geq 1}$ along which $\lim_{k\to \infty} \frac{H_{\textrm{sum}}(T_k+1)}{T_k}=0$. 

For this subsequence, we have from Eqn.\ \eqref{cesaro_mean_lt}:
\begin{eqnarray*}
	2\frac{H_{\textrm{sum}}(T_k+1)}{T_k}+\frac{1}{T_k}\sum_{t=1}^{T_k}H_{\textrm{sum}}(t) \geq 4 + \frac{1}{T_k}\sum_{t=1}^{T_k}\mathbb{E}(S(t)).
\end{eqnarray*}
Taking $k \to \infty$, we conclude that 
\begin{eqnarray}\label{ces_lim2}
\limsup_{T\to \infty} \frac{1}{T}\sum_{t=1}^{T}H_{\textrm{sum}}(t) \geq 6.
\end{eqnarray}
\textbf{Case II: $\liminf_{T\to \infty} \frac{H_{\textrm{sum}}(T+1)}{T}=\alpha >0$: }
From the definition of $\liminf$, it follows that there exists a finite $T_0$ such that, for all $T \geq T_0$, we have 
\begin{eqnarray} \label{liminf_eqn}
\frac{H_{\textrm{sum}}(T+1)}{T} \geq \frac{\alpha}{2}. 	
\end{eqnarray}
Thus, for any $T \geq T_0$, we can write 
\begin{eqnarray*}
	\frac{1}{T}\sum_{t=1}^{T} H_{\textrm{sum}}(t) \geq \frac{1}{T}\sum_{t=T_0+1}^{T} H_{\textrm{sum}}(t) \stackrel{(a)}{\geq} \frac{\alpha}{2T}\sum_{t=T_0}^{T-1}t = \Omega(T).
\end{eqnarray*}
Hence, in this case, we have 
\begin{eqnarray*}\label{lim_cost1}
\limsup_{T \to \infty} \frac{1}{T}\sum_{t=1}^{T} H_{\textrm{sum}}(t)  = \infty.	
\end{eqnarray*}
Hence, from Eqns. \eqref{ces_lim2} and \eqref{lim_cost1}, we conclude that, in either case, we have 
\begin{eqnarray} \label{lim_cost2}
	\limsup_{T \to \infty} \frac{1}{T}\sum_{t=1}^{T} H_{\textrm{sum}}(t)  \geq 6. 
\end{eqnarray}
Thus, In the case when $N=2$, using the result of Proposition \ref{improved_LB}, the competitive ratio is lower bounded by 
\[ \eta \geq \frac{6}{2^2}=1.5.\]
\end{IEEEproof}

\subsection{Proof of Theorem \ref{ld_opt}}\label{ld_opt_proof}
\begin{IEEEproof}
	Let $i^* = \arg \min_i p_i$. Now, under the action of any arbitrary policy $\pi$, at any slot $t \geq k$ and for all $k \geq 1$, we have
	\begin{eqnarray} \label{ld_lb}
		\mathbb{P}^\pi(\max_i h_i(t) \geq k) \geq \mathbb{P}^\pi(h_{i^*}(t) \geq k) \stackrel{(a)}{\geq} (1-p_{\min})^k, 
\end{eqnarray}
where the inequality (a) follows from the fact that consecutive $k$ erasures just prior to time $t$ for the $i^*$\textsuperscript{th} user (which occurs with probability $(1-p_{\min})^k$) ensures that the age of the user at time $t$ is at least $k$. \\
Next, we analyze the large-deviation exponent under the action of the \textsf{CMA} policy. Using the union bound, we have 
\begin{eqnarray}\label{ub}
\mathbb{P}\big(\max_i h_i(t) \geq k\big)=\mathbb{P}(\bigcup_{i=1}^Nh_i(t)\geq k)\leq \sum_{i=1}^N\mathbb{P}(h_i(t) \geq k).
\end{eqnarray}
Now, for any user $i$, the event $h_i(t) \geq k$ occurs if and only if it has been at least $k$ slots since the $i$\textsuperscript{th} user received a packet successfully before time $t$. Define $p_{\max}\equiv \max_i p_i$ and $p_{\min}\equiv \min_i p_i$. Since the \textsf{CMA} policy transmits other users successfully exactly once between two consecutive successful transmission to the user $i$ (due to its round-robin nature), it follows that, during the last $k$ slots prior to time $t$, at most $N-1$ users have successfully received a packet. Thus, we have the following bound:
\begin{eqnarray*}
	\mathbb{P}(h_i(t) \geq k) &\leq& \sum_{j=0}^{N-1} \binom{k}{j}p_{\max}^{j}(1-p_{\min})^{k-j} \\
	&\leq&  \binom{k}{N-1}(1-p_{\min})^k \times \\
	&&  \frac{1-p_{\min}}{(p_{\max}+p_{\min})-1}\bigg(\big(\frac{p_{\max}}{1-p_{\min}}\big)^{N}-1\bigg)\\	
&\leq& c'(N, \bm{p}) k^N (1-p_{\min})^k,
\end{eqnarray*} 
where we have used the bound $\binom{k}{N-1}\leq \frac{k^N}{(N-1)!}$ and defined $c'(N, \bm{p})\equiv \frac{1-p_{\min}}{(N-1)!((p_{\max}+p_{\min})-1)}\bigg(\big(\frac{p_{\max}}{1-p_{\min}}\big)^{N}-1\bigg) $\footnote{In the case $p_{\max}/(1-p_{\min})=1$, we take $c'\equiv N/(N-1)!$.}. Combining the above bound with Eqns.\ \eqref{ub} and \eqref{ld_lb}, we conclude that the \textsf{CMA} policy is optimal for the problem \eqref{ld_problem} and 
\begin{eqnarray*}
	-\lim_{k\to \infty}\lim_{t\to \infty}\frac{1}{k}\log \mathbb{P}^{\textsf{CMA}}(\max_i h_i(t) \geq k) = - \log(1-\min_{i=1}^Np_{i}).
\end{eqnarray*}

\end{IEEEproof}

%




\ifCLASSOPTIONcaptionsoff
  \newpage
\fi



%


\bibliographystyle{IEEEtran}
\bibliography{bibmobility}
\begin{IEEEbiography}[{\includegraphics[width=1in,height=1.25in,clip,keepaspectratio]{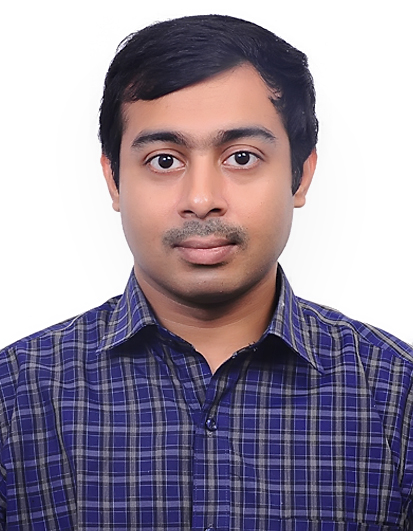}}]
{Abhishek Sinha}
is currently a faculty member in the School of Technology and Computer Science at the Tata Institute of Fundamental Research, Mumbai, India. Prior to joining TIFR, he had been with the Dept. of Electrical Engineering at the Indian Institute of Technology Madras as an Assistant Professor. He received his Ph.D. from the Massachusetts Institute of Technology, where he was affiliated with the Laboratory for Information and Decision Systems. Thereafter, Abhishek worked as a senior engineer at Qualcomm Research, San Diego, in the 5G standardization group. He obtained his M.E. degree in Telecommunication Engg. from the Indian Institute of Science, Bangalore, and B.E. degree in Electronics and Telecommunication Engg. from Jadavpur University, Kolkata, India. He is a recipient of the INSA Medal for Young Scientists (2021), Best Paper Awards in INFOCOM 2018 and MobiHoc 2016, and Jagadis Bose National Science Talent Search (JBNSTS) scholarship, Kolkata, India. His areas of interest include theoretical machine learning, networks, and information theory. 
\end{IEEEbiography}
\begin{IEEEbiography}
[{\includegraphics[width=1in,height=1.25in,clip,keepaspectratio]{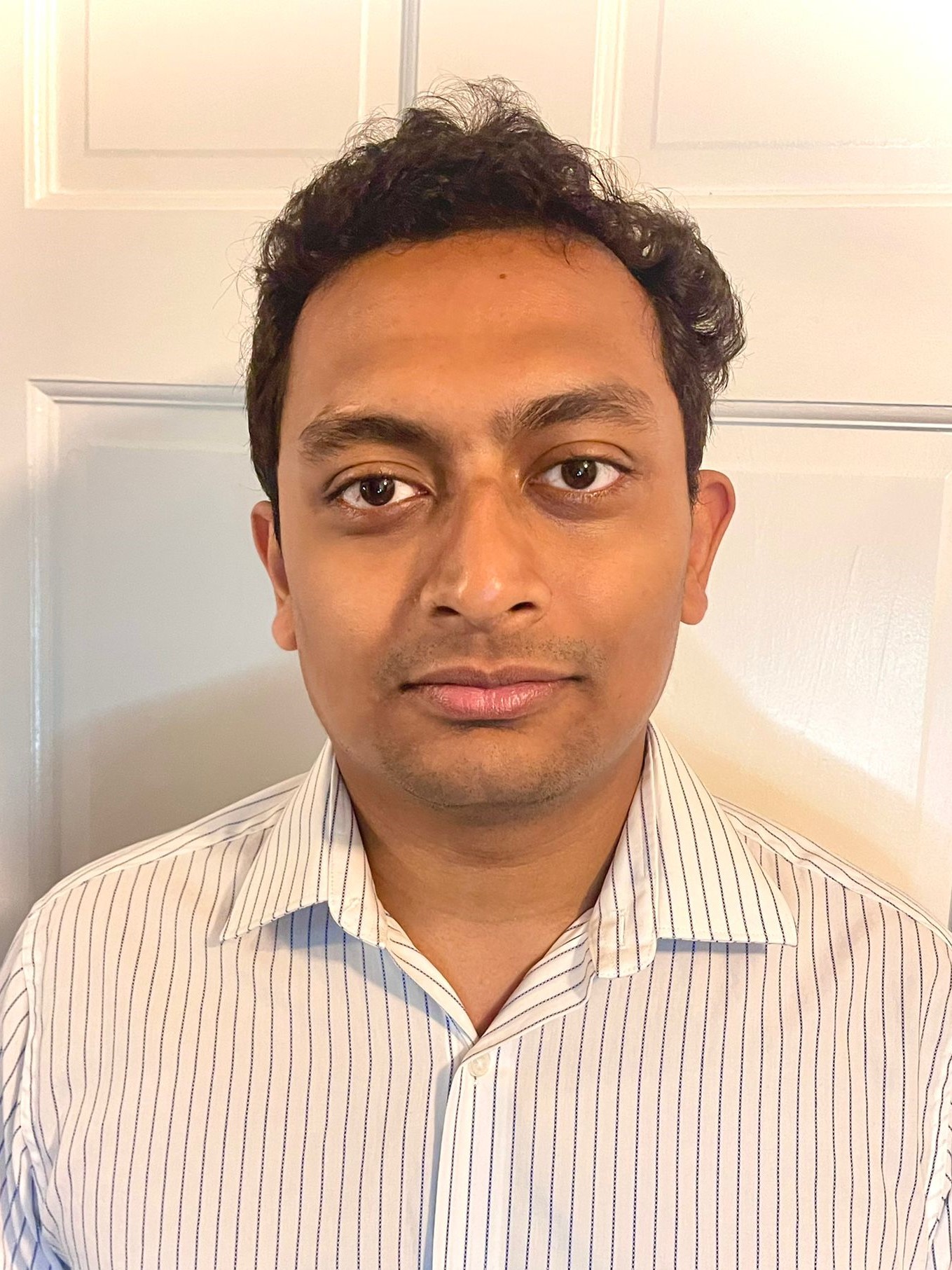}}]{Rajarshi Bhattacharjee}
{Rajarshi Bhattacharjee} received a B.E. degree in mechanical engineering from Jadavpur University, Kolkata, India, and an  M.Tech. degree in computer science from Indian Statistical Institute, Kolkata, India. He is currently pursuing a Ph.D. degree in computer science at the University of Massachusetts Amherst, USA. From 2019 to 2020, he was a research assistant at the Indian Institute of Technology Madras, India, under Prof. Abhishek Sinha, where he worked on online caching algorithms and on optimizing age-of-information in communication networks. His areas of interest include randomized algorithms, numerical linear algebra, online algorithms, and machine learning.
\end{IEEEbiography}

%







\end{document}